\documentclass[onecolumn,nofootinbib,12pt]{revtex4-1}
\usepackage[T1]{fontenc}
\usepackage[latin9]{inputenc}
\usepackage{verbatim}
\usepackage{amstext}
\usepackage{color}
\usepackage{graphicx,psfrag}
\usepackage{amssymb}
\makeindex
\def\eg{\emph{e.g.~}}
\def\ie{\emph{i.e.~}}

\def\beq{\begin{equation}}
\def\eeq{\end{equation}}
\def\bea{\begin{eqnarray}}
\def\eea{\end{eqnarray}}
\def\beg{\begin{lyxgreyedout}}
\def\eeg{\end{lyxgreyedout}}
\def\nn{\nonumber}

\begin{document}

\title{Introduction to the Quantum Theory of Elementary Cycles: \\ The Emergence of Space, Time and Quantum }\label{Dolce}

\author{Donatello Dolce}
\affiliation{University of Camerino, Piazza Cavour 19F, 62032 Camerino, Italy.}

\begin{abstract}
Elementary Cycles Theory is a self-consistent, unified formulation of quantum and relativistic physics. Here we introduce its basic quantum aspects. On one hand, Newton's law of inertia states that every isolated particle has persistent motion, \ie constant energy and momentum. On the other hand, the wave-particle duality associates  a space-time recurrence to the elementary particle energy-momentum.  Paraphrasing these two fundamental principles, Elementary Cycles Theory \emph{postulates} that every isolated elementary constituent of nature (every elementary particle) must be characterized by persistent intrinsic space-time periodicity.  Elementary particles are the elementary reference clocks of Nature. The space-time periodicity is determined by the kinematical state (energy and momentum), so that interactions imply modulations, and every system is decomposable in terms of \emph{modulated} elementary cycles. Undulatory mechanics is imposed as constraint ``overdetermining'' relativistic mechanics, similarly to Einstein's proposal of unification. Surprisingly this mathematically proves that the unification of quantum and relativistic physics is fully achieved by imposing an intrinsically cyclic (or compact) nature for relativistic space-time coordinates. In particular the Minkowskian time must be cyclic. The resulting classical mechanics are in fact fully consistent with relativity and  reproduces all the fundamental aspects of quantum-relativistic mechanics without explicit quantization. 
%Relativity only fixes the differential structure of space-time without giving prescriptions about the boundary. The constraint of covariant periodicity (or similar covariant boundary conditions) is fully consistent with the variational principle of relativistic theories.
This ``overdetermination'' just enforces both the local nature of relativistic space-time and the wave-particle duality. 
 Besides the unified description of relativistic and quantum dynamics, Elementary Cycles Theory implies  a fully geometrodynamical formulation of gauge interactions which, similarly to gravity and general relativity, is inferred as modulations of the elementary space-time clocks. This brings novel elements to address most of the fundamental open problems of modern physics. 
\end{abstract}%\markright{Customized Running Head for Odd Page} % default is Chapter Title.

\maketitle

\newpage

\setcounter{tocdepth}{4}
\tableofcontents

%\newpage

\begin{flushright}
		\vspace{2cm}
	\emph{\large To the memory of my father and my brother.}
			\vspace{2cm}
\end{flushright}

\section{Introduction}\label{ra_sec1}

\textit{Elementary Cycles Theory} (ECT) is a fully consistent formulation of \textit{Quantum Mechanics} (QM), relativistic physics, \textit{Gauge interactions} and other aspects of modern physics, obtained by postulating an intrinsically cyclic nature for Minkowskian space-time. As proven in many previous peer-reviewed papers, \eg \cite{Dolce:cycles,Dolce:ECunif,Dolce:FPIpaths,Dolce:TM2012,Dolce:Dice2012,Dolce:cycle,Dolce:ICHEP2012,Dolce:tune,Dolce:AdSCFT,Dolce:FQXi,Dolce:Dice,Dolce:2010ij,Dolce:2010zz,Dolce:2009ce,Dolce:EPJP,Dolce:SuperC,Dolce:Dice2014}, \emph{it provides an unprecedented, complete, exact, unified description of quantum and relativistic physics}. 

The idea is based on the empirical fact discovered by \textit{Louis de Broglie}  \cite{Broglie:1924,Broglie:1925}, knows as \textit{wave-particle duality}: elementary particles, \ie the elementary constituents of nature, have intrinsic recurrences (periodicities) in time and space. These are determined by the particle energy and momentum through the \textit{Planck constant} $h = 2 \pi \hbar$. All in all, this implicitly means that every system in nature can be consistently described in terms of elementary space-time cycles.

Paraphrasing Newton's first law (a \emph{free} particle has constant energy-momentum if viewed from an inertial frame) and de Broglie wave-particle duality originally formulated in terms of ``periodic phenomena''  \cite{Broglie:1924,Broglie:1925} (duality between energy-momentum and space-time quantum recurrence), we prove that classical-relativistic physics can be quantized by simply postulating that: 
\begin{quote}
Every free elementary particle, observed from an inertial frame, is an intrinsically ``periodic phenomenon'' whose persistent spatial and temporal recurrences are determined through the Planck constant by its persistent momentum and energy, respectively.  
\end{quote}
%These recurrences represent the quantization conditions of the relativistic physics. 

The novel aspect with respect to ordinary undulatory mechanics is that these recurrences must be imposed as \emph{constraints}, \ie by means of covariant Periodic Boundary Conditions (PBCs),  ``overdetermining'' the relativistic dynamics --- it can be imagined as a sort of relativistic generalization of the quantization of \textit{particle in a box} in which the Boundary Conditions (BCs) quantize the system. 

As we will prove, the resulting cyclic relativistic mechanics are formally equivalent to QM, even in its most advanced aspects. This equivalence is proven independently for all the formulations of QM:  the canonical formulation (axioms of QM), the Feynman path integral formulation, the Dirac formulation and second quantization; as well as for all the basic phenomenology of QM including all the textbook problems of QM, superconductivity, graphene physics, QED, AdS/CFT correspondence, holography,and so on  \cite{Dolce:cycles,Dolce:ECunif,Dolce:FPIpaths,Dolce:SuperC,Dolce:TM2012,Dolce:Dice2012,Dolce:cycle,Dolce:ICHEP2012,Dolce:tune,Dolce:AdSCFT,Dolce:FQXi,Dolce:Dice,Dolce:2010ij,Dolce:2010zz,Dolce:2009ce,Dolce:EPJP,Dolce:SuperC,Dolce:Dice2014}. The list of correspondences is  to long to be fully reported here. Its is a radically new view of quantum world. The reader must be ready to drastically reconsider, as consequence of the incontrovertible mathematical demonstrations, most of the academic interpretations of QM.   

In ECT every particle is described as an elementary relativistic clock  \cite{Dolce:FQXi,Dolce:tune}, in analogy with \textit{de Broglie's conjecture of the internal clock of particles}; see also the de Broglie ``periodic phenomenon''  \cite{1996FoPhL}, and (for Dirac particles) with the \textit{zitterbewegung} proposed by \textit{E. Schr\"odinger}. A similar description of elementary particles as clocks has also been recently adopted by \textit{Roger Penrose} in his book ``Cycles of Time: An Extraordinary New View of the Universe''  \cite{Penrose:cycles} and indirectly investigated by P. Cattillon and H. M\"uller experimental group  \cite{2008FoPh...38..659C,Lan01022013}, with emphasis on general relativity. 

The idea will be introduced by noticing that ECT has remarkable historical justifications in the ideas of some of the founding fathers of QM (de Broglie, Bohr, Planck, Einstein, Schr\"odinger, Feynman, Fermi, Sommerfeld, Dirac, Klein, Kaluza, etc; and more recently 't Hooft, Wilczek, Weinberg, etc).\footnote{Among the original motivations of ECT we have can also mention 't Hooft's Cellular Automata  \cite{Hooft:2014kka,hooft-2009-0,'tHooft:2007xi,'tHooft:2006sy,'tHooft:2001ar,'tHooft:1998fa}. 't Hooft  has proven that the there is a close relationship between the cyclic temporal dynamics of a particle moving on a circle (``continuous cogwheel model'', the analogous of a particle in a periodic time box) and the time evolution of the harmonic quantum oscillator.} 

Notably, it seems to realize fundamental aspects of  Einstein's original proposal to unify quantum and relativistic mechanics by means of relativistic constraints (represented by the space-time covariant PBCs of ECT) ``overdetermining'' the relativistic differential equations. Furthermore it can be regarded as an evolution (\ie  a generalization applicable to relativistic systems) of the \textit{Bohr-Sommerfeld quantization} or the \textit{WKB} method to solve quantum-relativistic problems. 

ECT has important confirmations in modern theoretical physics. It implies a geometrodynamical description of \textit{gauge invariance},    in perfect analogy with the geometrodynamical description of gravitational interaction in general relativity, according to Weyl's original proposal, \cite{Dolce:tune,Dolce:AdSCFT,Dolce:cycles,Dolce:cycle}. It also leads to the unprecedented mathematical demonstration of the validity of the so called Anti de Sitter / Conformal Field Theory (AdS/CFT) correspondence  \cite{Dolce:AdSCFT,Dolce:ICHEP2012,Dolce:2009ce}. Finally ECT  has deep relationships with ordinary String Theory and extra-dimensional theories (though ECT is purely four-dimensional), see  \cite{Dolce:tune,Dolce:cycles,Dolce:cycle} and forthcoming papers  \cite{Dolce:FPIpaths}. 

The unification of quantum and relativistic mechanics is based on \emph{unattackable mathematical proofs}, so it deserves the highest consideration from the physics community. %The reader must be ready to accept, as a consequence of the incontrovertible scientific arguments reported here, to reconsider most of the interpretations of quantum phenomena debated in literature. 
Its validity has been also successfully tested in non-trivial phenomenological aspects of condensed matter physics. For instance, the simple postulate of ECT has originated a novel, intuitive derivation of \textit{superconductivity} and related phenomenology, as well as of the electric properties of  \textit{carbon nanotubes}  \cite{Dolce:SuperC,Dolce:EPJP,Dolce:Dice2014}.  

By construction ECT does not involves hidden variables. Thus, ECT formulation of QM is not ruled out by Bell's or similar no-go theorems based on hidden variables. ECT proves that QM emerges as the statistical description of the deterministic ultra-fast cyclic dynamics associated to particles' internal clocks.  

The cyclic character of time proposed in ECT offers the possibility of a relational, emergent description of the time flow (with some analogies  to the ideas of R. Kastner and C. Rovelli   \cite{Kastner,Dolce:TM2012}). It clarifies the physical notion of time by clearly postulating that elementary particles are the elementary clocks of Nature. In other words, the Minkowskian time must have an intrinsically cyclic nature. The idea has been awarded by the IV prize in the 2011 \textit{FQXi} contest with the essay ``Clockwork quantum universe''  \cite{Dolce:FQXi}.

In the intention of the author, and thanks to the straightforwardness of the theory,   \emph{this paper is written in such a way to be adopted as a  general introduction to QM,  useful for students and non-experts on the subject}.  Indeed ECT  offers a self-consistent, extremely simple (``but not simpler'' A. Einstein) formulation of QM, based on intuitive, natural physical principles: every system in nature is describe by elementary space-time cycles. The hidden nature of quantum physics is nothing but pure relativistic harmony. 

The paper is structured in the following way: we will report de Broglie's undulatory mechanics by introducing a relativistic formalism; we will define ECT by means of the postulate of intrinsic periodicity, reporting historical motivations for such an assumption (de Broglie, Einstein, Penrose, 't Hooft, etc); we will derive, through simple but rigorous mathematics, the axioms of QM and the Feynman path integral directly from our  postulate of intrinsic periodicity; we will give examples of its applications in the solution of  textbook problems of QM and quantum phenomenology; finally we will mention some advanced applications in modern physics and conceptual implications.

\section{Overview}
\subsection{The concept of phase harmony and the basic formalism of undulatory mechanics}

ECT is based on the undulatory mechanics  and the related concept of phase harmony. Here we essentially report the wave-particle duality as originally described in de Broglie's PhD thesis  \cite{Broglie:1924,Broglie:1925}. It is however convenient to introduce it in a covariant way. This will constitute the basic formalism of ECT.
  
On one hand, according to undulatory mechanics, the \textit{Planck constant} relates the energy $E$ and the momentum $\vec p $ to the time periodicity $T$ and the wave-length (spatial periodicity) $\vec \lambda$, in the so called de Broglie-Planck relations
\begin{equation}
T = \frac{h}{E} ~, ~~~~~~~ \lambda_i = \frac{h}{p_i} \label{deBoglie:relations}~,(i=1, 2, 3)\,.
\end{equation} 

On the other hand, relativity relates, through the speed of light $ c $, the mass $ M $ to the rest energy $E(0)$, according to the universally known relation $E(0) = M c^2$. Hence, the combination of undulatory mechanics and relativity implies that to the particle mass must be associated a rest time periodicity, namely the Compton periodicity,  
\begin{equation}\label{Compt:time}
T_C = \frac{h}{ M c^2}\,. 
\end{equation}

In relativity the energy-momentum of the particle in a given reference frame is derivable from its mass by means of \textit{Lorentz transformations} \begin{equation}M ~~~~~~~\rightarrow~~~~~~~ p_\mu = \left \{\frac{E}{c} , - \vec p\right \} \end{equation} where $ E = \gamma M c^2 $ and $ \vec p = M c \gamma \vec \beta = \{p_1, p_2, p_3\}$; whereas $\gamma$ and $\beta$ are the Lorentz factors.  

It is convenient to write the temporal and spatial recurrences in a contravariant (space-like, tangent) \textit{four-vector} \begin{equation}T^\mu = \left \{T ,  \frac{\vec \lambda}{  c}\right \}  \,.\end{equation} %We will address it as \textit{space-time periodicity} 

Now we prove that $T^\mu$  is actually a four-vector that we will address as four-periodicity or space-time periodicity $T^\mu$. In analogy with the four-momentum, as proven by both de Broglie and Einstein (but rarely cited in literature), the space-time periodicity  is derivable  from the Compton periodicity  by means of Lorentz transformations: $T_C \rightarrow T^\mu$ such that   
\begin{equation}\label{Lorentz:Tc}
T_C c =  \gamma T c - \gamma \vec \beta \cdot \vec \lambda\,.   
\end{equation} 
Of course, the phase of relativistic (matter) waves $e^{-i p_\mu x^\mu / \hbar} $ must be an invariant: $  M c^2 \tau =  p_\mu x^\mu = invariat $; where $\tau$ is the proper-time and $x^\mu = \{ c t , \vec x \}$.  
By substituting  $T_C$ in the Compton relation (\ref{Compt:time})  we obtain the invariant relation, also known as de Broglie \emph{phase harmony relation} \begin{equation}  M c^2 T_C  =  \gamma M c^2 T -  M c \gamma \vec \beta \cdot \vec \lambda=  E T - \vec p \cdot \vec \lambda = c p_\mu T^\mu =  h\,. \end{equation}

The energy-momentum and the space-time periodicity of the particle in a generic reference frame --- similarly to the mass and the Compton period for the rest particle --- are therefore determined by the covariant phae-harmony relation \footnote{The properties of the four-vector $T^\mu$ are described in dept in  \cite{Dolce:tune}. It can be generalized to the interaction case where it must vary locally (instantaneous four-periodicity). In general, it turns out to be a tangent four-vector (it transforms as $d x^\mu$) so it makes sense to write the phase harmony in the following way $T^\mu = h / c p^\mu$. This is the relativistic generalization, rarely reported in literature, of the  de Broglie -Planck relation.}
\begin{equation}  M c^2 T_C  = h ~~~~~~~\rightarrow ~~~~~~~ c p_\mu T^\mu =  h\,.\label{de:Broglie:phase:harmony} \end{equation} 

The phase harmony implies that \emph{the momentum-energy and the space-time periodicity are dual quantities} --- they are two faces of the same coin. This is the basic meaning of undulatory mechanics and of the wave-particle duality expressed by (\ref{deBoglie:relations}). For instance, as it can be easily checked dividing by $\hbar^2$, the relativistic dispersion relation of the energy-momentum can be equivalently expressed in terms of the relativistic dispersion relation of the space-time periodicity 
\begin{equation}\label{disp:relat}
M^2 c^4 = p_\mu p^\mu = E^2 - \vec p^2 c^2 ~~~~~~~\mathrel{\mathop{\longleftrightarrow}^\hbar} ~~~~~~~ \frac{1}{T^2_c} = \frac{1}{T^\mu} \frac{1}{T_\mu} = \frac{1}{T^2} - \sum_{i=1}^3 \frac{c^2}{(\lambda^i)^2}\,.
\end{equation}

This means that the temporal period $T$ is globally modulated under transformations of the free particle reference frames according to the relativistic Doppler effect. 

\subsection{Postulate of elementary space-time cycles}
 
Classical-relativistic mechanics is based on simple postulates, such as Newton's laws and Einstein's principle of equivalence, with universally accepted physical meaning. The other pillar of modern physics, \ie QM, has however an axiomatic formulation (see the  axioms derived from EC in this paper) whose physical meaning is still largely debated by experts on the field. Furthermore, it is notoriously problematic to conciliate together relativity and QM. It is not a case that some of the most eminent physicists (\eg Einstein, Feynman, or more recently 't Hooft, Weinberg, Wilczek) have expressed the necessity of a deeper comprehension of QM, as well as possible new physics beyond QM (unfortunately many other physicists seem to have given up with the effort to understand quantum world in terms of simple, physical, first principles, and quantum weirdnesses are sometimes accepted as dogmas). 

The novel fascinating physical interpretation of QM emerges by combining together Newton's law of inertia and the undulatory mechanics of elementary particles described above. From these, as already pointed out, it necessarily follows that elementary particles of persistent energy and momentum, \ie free particles, must have persistent recurrences in time and space. 

Of course Newton's law of inertia describes an ideal case. In everyday life we are used to bodies that have not uniform motion because it is difficult to isolate bodies from interactions (local variations of energy-momentum). We are used to bodies that come to rest because of friction (see concept of inertia in Aristotle). 

Similarly, as we will discuss, the intrinsic periodicity proposed in ECT (implicit in the wave-particle duality) is an ideal condition typical of pure quantum systems: \eg free relativistic particles, but also high energies, small distances, or very low temperatures. In everyday life, of course, we do not experience a perfectly periodic world (must be said, however, that the universe is full of effective periodic phenomena). 

Ordinary objects are composed by a large number of elementary particles (consider that, even neglecting interactions, a simple set of two persistent ``periodic phenomena'' forms an ergodic system); the quantum recurrences are typically very small  \cite{Dolce:2009ce,Dolce:FQXi} (the Planck constant is so ``small'' that it can be neglected in ordinary world); we must take into account interactions which correspond to local modulations of the space-time recurrences  \cite{Dolce:cycles,Dolce:tune,Dolce:AdSCFT} (even very complex and chaotic evolution can be always regarded as interacting elementary ``periodic phenomena'', as implicit in QFT); we have the thermal noise implying chaotic interactions (the Brownian motion tends to destroy the de Broglie-Planck periodicity, \ie the perfect coherence of pure quantum systems  \cite{Dolce:SuperC,Dolce:EPJP,Dolce:Dice2014}). Nevertheless, as we will argue, the ideal case of elementary particles perfect recurrences describes the quantum world with all its peculiar phenomena whereas the classical limit is obtained when the recurrences can be neglected in the effective description.

In ECT we have the rare privilege to introduce a simple postulate in physics, in addition to the ordinary postulates of classical-relativistic mechanics. This is the postulate of intrinsic periodicity (imposed as constraint but implicit in Newton's and de Broglie's principles) from which the whole construction of the theory is derived. 
In particular, in ECT it is not necessary to assume the axioms of QM or any other quantization; these are inferred directly from the postulate. Intrinsic periodicity represents the quantization condition for the classical-relativistic mechanics and no further quantization conditions are required. Intrinsic periodicity is, above any doubt  \cite{Dolce:cycles,Dolce:ECunif,Dolce:FPIpaths,Dolce:TM2012,Dolce:Dice2012,Dolce:cycle,Dolce:ICHEP2012,Dolce:tune,Dolce:AdSCFT,Dolce:FQXi,Dolce:Dice,Dolce:2010ij,Dolce:2010zz,Dolce:2009ce}, the missing link between relativistic and quantum physics: the mathematics proving it is unattackable, so it is a duty to try to figure out about this revolutionary simple formulation of quantum-relativistic mechanics. The quantum world is pure relativistic harmony rather then an obscure, paradoxical, ``weird'', axiomatic theory. 

The postulate of intrinsic periodicity (elementary space-time cycles) in ECT can be  enunciated in three equivalent ways:

\vspace{0.5cm}
\hspace{-0.5cm}
\fbox{ \hspace{-0.8cm}\parbox[1em]{\textwidth}
{\begin{itemize}\label{postulate}
  \item An elementary particle of mass $M$ is an intrinsically ``periodic phenomenon'' of Compton periodicity $T_C = h / M c^2$.
%\end{itemize}
%\begin{itemize}
\item An elementary particle of mass $M$ is a reference elementary clock of Compton periodicity $T_C = h / M c^2$.
\item An elementary particle of mass $M$ is a vibrating string of fundamental Compton periodicity $T_C = h / M c^2$.
\end{itemize}
}}

\vspace{0.5cm}
This also defines the concept of  ``elementary particle'' in ECT.  A rest elementary particle is any ontological entity described  as a persistent periodic phenomenon (or, in case of composite, not fundamental particles are entities which can be approximated as periodic phenomena in the effective description). We will use the term ``periodic phenomenon'', reference clock or vibrating string as synonymous  of elementary quantum particle.

In terms of the wave-particle duality, \ie of the relationship between Compton periodicity and rest mass allowed by the Planck constant, the classical-relativistic counterpart of this postulate is that a particle at rest has a  mass determined by the Compton periodicity, $M = h / T_C c^2$. 

According to the covariant description of undulatory mechanics given in the previous section, the postulate of elementary cycles can be easily generalized to isolated (free) particles viewed from generic inertial reference frames 

\vspace{0.5cm}
\hspace{-0.5cm}
\fbox{ \hspace{-0.8cm}\parbox[1em]{\textwidth}
{ \begin{itemize}
\item A free elementary particle of four-momentum $p_\mu$, observed from an inertial reference frame, is an intrinsic ``periodic phenomenon''/reference clock/vibrating string of persistent de Broglie-Planck space-time periodicity $T^\mu$, according to the  phase harmony relation $c p_\mu T^\mu = h$. 
\end{itemize} }}
\vspace{0.5cm}

The classical-relativistic counterpart of this principle is that every isolated relativistic particle has constant energy-momentum. That is, according to Newton, every isolated relativistic particle moves in uniform rectilinear motion, as long as it does not interact. In fact, by means of the Planck constant, a periodic phenomenon with persistent space-time periodicity $T^\mu$ has constant energy-momentum $p_\mu$, see phase harmony relation (\ref{de:Broglie:phase:harmony}) or (\ref{deBoglie:relations}). 
 
 \subsubsection{``Periodic phenomena''}
 
The first enunciation of the postulate of intrinsic periodicity pays tribute to de Broglie. In his seminal PhD thesis  \cite{Broglie:1924,Broglie:1925} he in fact stated the hypothesis of wave-particle duality in this form:

\begin{quote}
\emph{``To each isolated parcel of energy [elementary particle] with a proper mass $M$, one may associate a periodic phenomenon of [Compton] periodicity $T_C = h / M c^2 $. The [Compton] periodicity is to be measured, of course, in the rest frame of the particle. This hypothesis is the basis of our theory: it is worth as much, like all hypotheses, as can be deduced from its consequences.''} L. de Broglie (1924).
 \end{quote}
 
... and we will see that these consequences go incredibly far, much farther than the applications commonly attributed to the wave-particle duality. 
The difference with respect to ordinary applications of the wave-particle duality is that the postulate of elementary cycles enforces the \textit{wave-particle duality}, which must be encoded directly into the structure of relativistic space-time. It brings \textit{de Broglie}'s hypothesis of wave-particle duality to its extreme consequences, as relativity brings  the hypothesis of constant speed of light to its extreme consequences. The intrinsic recurrences will be imposed as constraints ``overdetermining'' relativistic space-time dynamics. The resulting cyclic relativistic dynamics will be formally equivalent, for all the fundamental aspects, to the ordinary quantum-relativistic mechanics in all its formulations.

\subsubsection{The elementary reference clocks of Nature}

The second way to postulate the principle of ECT is also indirectly suggested by de Broglie, but also by Compton, Einstein and Penrose. All these authors pointed out that the phase harmony described above (wave-particle duality), which associates to every particle a periodic phenomenon, implies that every particle must be regarded as having an ``internal clock'' of rest periodicity $T_C$. 

The fact that a periodic phenomenon of persistent periodicity, \ie an elementary particle, can be regarded as an elementary clock follows clearly from Einstein's definition of relativistic clock  \cite{Einstein:1910}:

\begin{quote} \emph{``By a clock we understand anything characterized by a  phenomenon passing periodically through identical phases so that we must  assume, by the principle of sufficient reason, that all that happens in a  given period is identical with all that happens in an arbitrary period''}, A. Einstein (1910).
\end{quote}

Such a description of particles as elementary clocks  has been, for instance, recently adopted by R. Penrose which, in agreement with our postulate, wrote 

\begin{quote}
\emph{``for there is a clear sense in which any individual (stable) [isolated and at rest] massive particle plays a role as a virtually perfect clock. [...] In other words, any stable massive particle behaves as a very precise quantum clock, which ticks away with [Compton periodicity].''} R. Penrose (2011)
\end{quote}

Massless particles have infinite Compton periodicity. In principle they can have arbitrary large space-time periodicities, \ie arbitrary small energy-momentum (zero rest energy). We anticipate that massless particles like photons play a central role in ECT. They define the emphatically non-compact space-time typical of the ordinary interpretation of relativity  \cite{Dolce:cycles,Dolce:tune,Dolce:AdSCFT}. That is the reference space-time for the relational description of the elementary cycles. 

Here we must consider that mass is crucial for a consistent description of time flow in ECT. We may say that massless particles, due to their infinite Compton periodicity, have ``frozen'' internal clocks.  As noted by R. Penrose: 

\begin{quote}
\emph{``Massless particles (\eg photons or gravitons) alone cannot be used to make a clock, because their frequencies would have to be zero; a photon would take until eternity before its internal 'clock' gets even to its first 'tick'! To put this another way, it would appear that rest-mass is a necessary ingredient for the building of a clock, so if eventually there is little around which has any rest-mass, the capacity for making measurements of the passage of time would be lost''} R. Penrose (2011). 
\end{quote}

That is to say, ordinary relativistic space-time --- Minkowskian space and time are intimately mixed --- by itself does not provide sufficient elements  to fully understand the flow of time of a physical systems. The notion of time is the greatest mystery of physics. ECT clearly defines the elementary clocks of Nature. A consistent description of the relativistic time flow is only possible if some elementary particles acquire mass, \ie a finite Compton periodicity, otherwise all the particles would be on the light-cone with ``frozen'' internal clocks. Nature defines  elementary clocks through particles' masses. This often neglected element, in addition to the Minkowskian time coordinate, is crucial to understand the notion of time in physics and the meaning of the arrow of time, see Sec.(\ref{timeflow}) and citations thereby. %\footnote{Without the definition of the elementary clocks of Nature, one may be tempted, as some authors, to rejected the notion of time in physics, \eg J. Barbour. Physics is an experimental science and the statement that time does not exist, though we do not completely understand it, must be considered a failure of science.}    

We generally speak about space-time cycles, however the distinctive character of ECT is essentially given by the periodicity in time, rather than in space. If fact, for a massive particle at rest, $\vec p \rightarrow 0$, the spatial periodicity tends to infinite (infinite wave-length); only the periodicity in time si left, which coincides with the Compton periodicity. The whole ECT can be equivalently derived by simply assuming periodicity in the Minkowskain time, the periodicity in space is a consequence of the equations of motion, see the original formulation based on compact time in \cite{Dolce:2009ce}.

\subsubsection{Vibrating strings}

The term  ``periodic phenomenon'' used in the first enunciation of the postulate may seem to be vague - or it may be erroneously confused with the ``matter-wave'' of the ordinary semi-classical formulation of QM. Its meaning is clarified by the third enunciation of the postulate: an intrinsic 	``periodic phenomenon'' of given periodicity is a \emph{vibrating string, \ie the set of all the harmonic eigenmodes allowed by its fundamental vibrational period}, \cite{Dolce:cycles,Dolce:FPIpaths,Dolce:ECunif}.

 In general, by means of discrete Fourier transform, a periodic phenomenon of finite period  (\eg a periodic function) can be always represented as the sum of harmonic eigenmodes, see Fig.(\ref{harmonics}).  The difference with ordinary undulatory mechanics is thus that  \textit{we impose intrinsic periodicity as constraint to ``overdetermine'' the relativistic dynamics of the free elementary particle}. 
 
 More precisely, a free quantum particle of persistent four momentum $p_\mu$, and its quantum excitations, must be described as vibrations of space-time associated to its fundamental four-periodicity $T^\mu$, according to the relativistic phase harmony. We will see how to derive an ordinary quantum field from this. Here we  only mention that this fundamental vibrating string describes the elementary quantum oscillator at the base of ordinary quantum fields.

 In the semi-classical formulation of QM or in ordinary field theory such  ``periodic phenomenon'' associated to the particle is typically limited to the monochromatic wave of corresponding angular frequency $\omega = 2 \pi / T$ and angular wave-number $k_i = 2 \pi / \lambda^i$. The ``matter-wave'' is then typically quantized by imposing commutation relations in order to obtain the quantum behavior of the free particle.  Its (normally ordered) quantized energy spectrum is in fact $E_n = n \hbar \omega$ (\ie the energy spectrum of the normally ordered quantum harmonic oscillator).  This method is at the base of the so-called second quantization of ordinary  Quantum Field Theory (QFT). 
 
 The central point is that, in ECT,  the most general periodic phenomenon of periodicity $T = 2 \pi / \omega$ is not a simple monochromatic wave. It is given by all its  possible harmonics, \ie by all the wave components with discretized angular frequencies  $\omega_n = n \omega$, Fig.(\ref{harmonics}). For instance, we may notice that, through the Planck constant (without introducing any further quantization condition), the vibrating string of time period $T$ provides the same energy spectrum of the free particle $E_n = \hbar \omega_n = n \hbar \omega$ prescribed by ordinary (normally ordered) QFT, see also the Planck energy spectrum.  As in ordinary strings, we will see that the elementary particles will be described in terms vibrations associated to compact dimensions. 
  
   In general, the higher harmonics describe the quantum excitations of the particles (\eg multiparticle states) whereas the fundamental harmonic is the de Broglie \textit{matter wave} of ordinary undulatory mechanics or field theory --- the negative vibrational modes correspond to antimatter (``anti-particles''). Notice again that the hypothesis of pure, \emph{persistent} periodicity is limited to free, isolated quantum-relativistic particles (principle of inertia). According to the phase harmony, interactions, \ie local variations of the local energy-momentum of the particle, imply corresponding local modulations of space-time periodicity and thus local deformations of the corresponding harmonic set --- similarly to a non-homogeneous string. 
 
 The classical-particle limit is naturally obtained from the non-relativistic limit of this pure ``periodic phenomenon''. As we will see, in the non-relativistic limit the rest energy $E(0) = M c^2$ can be ``neglected'' (it forms an infinite energy gap). That is, in the non-relativistic limit the Compton periodicity $T_C$ tends to zero. It can be neglected as well, and the quantization associated to the  ``periodic phenomenon'' is lost. So, a free non-relativistic particle is effectively described by  the continuous energy and momentum spectra. In this limit we actually get a corpuscular description.
 
 For interacting non-relativistic particles (\eg bounded in a potential), the BCs quantizing the systems, and the corresponding harmonic sets turn out to be determined by the geometry or by the potential (\eg as for a particle in a box or the atomic orbitals). Intrinsic periodicity in this case corresponds to the requirement that only closed space-time orbits (with integer number of recurrences) are allowed, in analogy with the Bohr-Sommerfeld quantization, see sec.(\ref{Bohr-Sommerfeld}).  %Similar to the semi-classical problem of a particle in a box or a vibrating string, the constraint of intrinsic periodicity implies that the a relativistic particle can be regarded as an harmonic system, \ie a wave packet of all the possible harmonics eigenmodes associated to its de Broglie-Planck periodicity. 
 
 ECT indeed supports a stringy description of elementary particles. We conclude this digression by reporting that ECT inherits fundamental properties of ordinary String Theory. ECT is essentially a string theory defined on a ``compact world-line'' rather then on a compact ``world-sheet''  \cite{Dolce:cycles,Dolce:SuperC,Dolce:TM2012,Dolce:Dice2012,Dolce:cycle,Dolce:ICHEP2012,Dolce:tune,Dolce:AdSCFT,Dolce:FQXi,Dolce:Dice,Dolce:2010ij,Dolce:2010zz,Dolce:2009ce}, see also the forthcoming paper  \cite{Dolce:FPIpaths}.
 
 \subsubsection{Einstein's ``overdetermination'' of relativistic mechanics and ``supercausality''}
 
As already said, in ETC  the constraint of intrinsic periodicity is imposed a constraint to quantize relativistic dynamics. 
 Interesting enough, the idea to ``overdetermine'' relativistic differential equations with covariant --- relativistic --- constraints in order to obtain in an unified view,  quantum dynamics out of relativistic ones was originally proposed by Einstein. As reported by A. Pais, Einstein was convinced that ``it is necessary to start from  classical [relativistic] field theories [\eg undulatory mechanics] and ask that quantum laws emerge from constraints imposed to them'' \cite{Pais:einstein}.  Einstein wrote \cite{Einstein:1923} 

\begin{quote}
\emph{For sure; [in order to describe quantum states] we must just overdetermine the variables of the [relativistic] matter wave or field by means of constraints. [...] The [relativistic] dynamics of the particles would be overdetermined in such a way that the initial conditions would be subject to restrictive constraints''.} A. Einstein (1923).     
\end{quote}

Then he added some requirements for these constraints such as covariance and compatibility with electromagnetism and gravity.  These requirements are actually fully satisfied by the constraint of contravariant intrinsic periodicity of ECT. Below we will discuss how covariant PBCs, or more in general Dirichlet or Neumann contravariant BCs, are the only natural constraints allowed by the variational principle for a free relativistic theory.  

Einstein considered the example of the atomic orbitals. Although this case concerns interaction it is easy to see that the atomic orbitals can be interpreted in terms of intrinsic periodicity. As we will argue rigorously in sec.(\ref{Coulomb}), similarly (but not identically, as we are not restricted to circular orbits) to Bohr's description, the atomic orbitals are the generalization to interactions of the harmonics of a (non-homogeneous) vibrating space-time string. In fact, in the atomic orbitals the wave function of the electron must be constrained to be periodic --- integer number of space-time recurrences --- along the Coulomb potential. The frequency spectrum, and thus the Bohr spectrum (or the quantization of the angular momentum), is determined by the condition that the wave function must be closed along its space-time orbits, similarly to the harmonics of a vibrating string. In other words principal quantum number $n$ of the atomic orbitals can be thought of as associated to the vibrational space-time modes of the locally deformed (according to the Coulomb potential) vibrating string whose rest periodicity is the Compton time of the electron (as we will see the other orbital quantum number result, due to the spherical symmetry, as the vibrational  modes of a spherical membrane, \ie the spherical harmonics). 

The term ``overdetermination'' used by Einstein means that these constraints encoding quantum dynamics into relativistic mechanics must be added to (or must replace) the ordinary initial (and final) boundary conditions of ordinary classical-relativistic theories, (\ie the so called stationary BCs of classical theories). For instance, for a free classical-relativistic particle, in order to satisfy the variational principle at the boundary, it is typically assumed null variations at the extremal times of its evolution (noticed that this  yields a tautology in even classical mechanics: how can the particle possibly knows at the beginning of its evolution the final time at which its trajectory will have null variation?). This also fixes the energy of the classical trajectory. 

In ECT, however, the stationary conditions are replaced by PBCs, which are determined by the energy-momentum of the particle, according to the phase harmony. For every given initial (local, in case of interactions) energy-momentum of the classical particle we must impose the corresponding (local) PBCs (in this way the tautology associated to the stationary BCs is solved). This ``overdetermination'' quantizes the system, see for instance the simple case of the Black-Body radiation in sec.(\ref{BlackBody}) and Fig.(\ref{harmonics}).

Einstein's proposal of ``overdetermination'' indeed allows us to introduce some advanced mathematical aspects of ECT --- though these are not relevant for the pedagogical scope of this paper. From a mathematical point of view, the hypothesis of intrinsic space-time periodicity is implemented by imposing a compactification length to the ordinary space-time coordinates and contraviariant PBCs. This formalism guarantees that the PBCs ``overdetermining'' the relativistic differential equations satisfy the variational principle of the corresponding relativistic actions --- exactly as the  stationary BCs of ordinary classical and quantum theories. Covariant PBCs or, more in general combinations of Dirichlet and Neumann BCs, are the only BCs allowed by the variational principles for a free relativistic particle,  \cite{Dolce:2009ce}. 

The compatibility with the variational principle guaranties that the covariant (local) PBCs preserve all the fundamental properties of the relativistic theory, \eg the covariance and causality.  Of course, new ---relativistic --- phenomena must emerges from these BCs, along with the purely relativistic ones. ECT proofs that the novel manifestation associated to these relativistic cyclic dynamics is exactly  ordinary quantum-relativistic dynamics in all its fundamental phenomenology and mathematics.

We have seen that in the free case the space-time period $T^\mu$ transforms in a contravariant way. Indeed, it is easy to show that these PBCs imposed as constraint to a bosonic relativistic theory satisfy the variational principle \cite{Dolce:2009ce,Dolce:tune}. Actually, ECT has a full compatibility with both special and general relativity. This is because relativity only fixes the differential structure of space-time without concerning about the boundaries. For instance, once that the rate of a clock  is fixed in a given reference system (\eg the rest frame), relativity only prescribes how this rate varies in the other reference frames (relativistic Doppler effect). Nevertheless in relativistic-physics it is necessary to introduce some first principle to define the elementary clocks of Nature. The postulate of intrinsic possibility satisfies this necessity.

Notice that in this paper we will only consider PBCs as we will not discuss spin, but different kinds of BCs (provided that they satisfy the variational principle) are allowed by different kinds of relativistic dynamics --- \eg Dirac or Klein-Gordon dynamics. Indeed bosonic or fermionics quantum dynamics are both characterized by different kind of BCs in compact space-time.  

 In physics as in mathematics, a well formulated problem needs differential equations as well as BCs (Cauchy problem), but relativity is not able to answer to the simple question: \textit{where is and what is fixed at the boundary of space-time}? The answer to this comes from QM. QM is about BCs as noticed by the fathers of quantum theory (de Broglie, Bohr, Sommerfeld, and so on). 
 
 Relativity only fixes, through the metric, the differential structure of space-time, \ie how the periodicities of the clock varies. Actually, special and general relativity was originally conceived by Einstein in terms of relativistic clocks modulations. However the rates  of the fundamental clocks of Nature (\ie the PBCs of relativistic dynamics) is fixed by QM (Compton periodicity) rather than by relativity, as clearly postulated in ECT. This analogy between particles, \ie elementary space-time cycles, and relativistic clocks reveals the full mathematical compatibility of ECT with special and general relativity. 
 
 This also allows us to understand causality in ECT \footnote{Contrarily to ECT, in theories based on static time periodicity, such as in the theory of Closed Timelike Curves, causality is manifestly violated}. The essential point to bear in mind here is that the space-time periodicity of the particle varies in a controvariant way with respect to its energy-momentum. That is, they are determined by the energy-momentum, through the local phase harmony condition (de Broglie-Planck relation). The energy propagates according to the retarded relativistic potential so that, when it is absorbed by a particle during interaction, the particle periodicity  (\ie together with the energy) is modulated in a local, retarded way  \cite{Dolce:2009ce,Dolce:tune,Dolce:AdSCFT,Dolce:TM2012,Dolce:FQXi}. We will give some more detail about this when we will discuss the time flow in ECT, Sec.(\ref{timeflow}).  Actually, by using Einstein's terminology \cite{Pais:einstein,Einstein:1923}, we can say that the constraint of intrinsic periodicity yields a ``supercausality''. Indeed, in ECT the local and causal nature of space-time is enforced, together with the wave-particle duality.  
 
 \subsubsection{Elementary space-time cycles}

From a formal point of view, the novel aspect introduced by ECT is that relativistic elementary particles must be described in terms of cyclic Minkowskian space-time coordinates (more in general compact space-time dimensions with different type of topologies, depending on the particle quantum number). The space-time periodicity (\ie the space-time compactification length) is determined, through  Lorentz transformations, by the Compton periodicity, \ie by the rest energy (mass).  In this way undulatory mechanics --- and, as we will see, the whole QM --- are directly encoded into the relativistic space-time geometrodynamics. 

We must bear in mind that every system in nature is totally composed by elementary particles, and every elementary particle is a periodic phenomenon (with local modulations determined by interactions). It immediately follows that every system in nature must be described in terms of elementary (modulated) space-time cycles. In ECT this possibility is realized by the fact that, according to the postulate of intrinsic periodicity, the space-time coordinates are treated as (dimensional) angular variables.  ECT defines relativistic differential problems by adding intrinsic covariant BCs to the relativistic space-time dynamics describing every single particle. 

 In undulatory mechanics this is implicit in the fact that space-time coordinates always appear in phasors or waves functions.  An angular variable is a variable which appears as the argument of a periodic phenomenon, for instance,  in a wave or a phasor, $e^{-i E t / \hbar}$: $t$ here is an angular (or periodic) variable of period $T = h / E$.  The success of undulatory mechanics and QFT essentially tells us that every system in Nature (every set of particles) is described by a set of ``periodic phenomena''  (\eg phasors or wave functions). Inevitably every physical system in nature must be described by a set (one for each elementary particle) of space-time periodic (angular) coordinates. In the case of PBCs,  this can be equivalently stated for a generic reference frame as follows: \textit{in elementary particles, the space-time dimensions have a relativistic cyclic nature, with local space-time periodicities $T^\mu$}.

Another formal consequence of the ECT is that, by generalizing the above arguments to a rest particle (Compton clock), \textit{ the proper-time parameter of every elementary particle has an intrinsically cyclic (compact) character of duration $T_C$}. Similarly to the cyclic (compact) space-time dimensions, the proper-time parameter of elementary particles must have PBCs (or other kind of BCs allowed by the variational principle for that relativistic field action). 

 For the conceptual consistency of this description  it is crucial to notice that massless particles, having infinite Compton periodicity, are associated to a proper-time parameter with infinite compactification length, \ie frozen internal clock as pointed out by Penrose. Hence,   massless particles can potentially have infinite compactification lengths for the space-time coordinates. This potentially infinite space-time associated to massless particles plays the role of the reference (non-compact) space-time, analogous to the (emphatically non-compact) space-time of ordinary relativity. This guarantees the correct relational description of physical systems. 
 
 Interactions, \ie local and retarded modulations of space-time periodicities, are therefore described by local and retarded deformations of the space-time compactification lengths. The corresponding geometrodynamics of the space-time boundary and of the metric actually describe in a unified way gauge and gravitational interactions, respectively \cite{Dolce:tune,Dolce:AdSCFT}, see also \cite{Dolce:cycles,Dolce:TM2012,Dolce:Dice2012,Dolce:cycle,Dolce:ICHEP2012,Dolce:FQXi,Dolce:Dice,Dolce:2010ij,Dolce:2010zz,Dolce:2009ce}. 

In ECT elementary particles themselves can be regarded as vibrations of space-time. Similarly to relativity, there is not an universal space-time. Each elementary particle defines its own cyclic space-time.  The space-time defined by massless particles is potentially non-compact and serves as reference for the massive particles space-times. Massless particles are actually the mediators of interactions.

Finally, for its beauty, we have to mention another fascinating view allowed by ECT. We have a perfect correspondence between the full relativistic generalization of the theory of sound and ECT. By full relativistic generalization of the theory of sound we do not only mean relativistic waves, but also the fact that the sound sources can vibrate in time and not only in space as ordinary sound. Thus, massive elementary particles, with they vibrations in space and time, are the relativistic sources of sound. In fact we have postulated that every elementary particle is a vibrating space-time string. 

Each massive particle has associated fundamental frequency established by the value of their masses through the Compton relation. Each massive particle of the Standard Model produces a peculiar relativistic ``note'', depending on its Compton time. These notes are not simple sinusoidal waves of corresponding frequencies, \ie these are not simply the ``matter wave'' of ordinary QM. The relativistic ``sound'' generated by elementary particles  has a ``timbre'', it is composed by all the possible frequencies allowed by their fundamental Compton periodicities. 

The ``medium'' on which this relativistic ``sound`` propagates is the space-time(s) associated to the fundamental massless particles of the Standard Model, \ie photons and gravitons. As the air in a concert hall, the cyclic space-time of the massless particles can potentially have infinite fundamental periodicities. This is the meaning the statement that, in ECT, massless particles provide the reference potentially non-compact space-time (since in ECT the elementary particles are space-time vibrations, the old ``ether'' can be regarded as the space-time generated by massless particles, potentially non-compact, vibrations). In other words they can transfer and resonate at all the possible notes, see Fig.(\ref{corponero}), allowing to the elementary relativistic sound sources (\ie to the elementary particles) to communicate each other (\ie to interact). As in music, time is defined in relation to the peach (frequency) of the notes. 

ECT reveals that the unified description of physics is essentially the full relativistic generalization of the theory of sound. The universe is pure relativistic harmony and, we should add, it could not be otherwise. 

Now, with a little patience, we can finally prove all this fascinating view in a rigorous mathematical way.

\section{Full Derivation of Quantum Mechanics from elementary space-time cycles}

In ETC the postulate of intrinsic periodicity  represents the quantization condition of the relativistic dynamics from which we now derive  all fundamental axioms of QM, as well as the Dirac quantization (commutation relations). %axiom of the state (Hilbert space), of the observables (Hamiltonian and Hilbert operators), of the motion (Schr\"odinger equation), of the commutation relations (and Heisenberg uncertain relation) and the axiom of the measurement  (Bohn principle).  
We shall also describe how the purely relativistic (cyclic) dynamics associated to the ECT leads to the ordinary Feynman path integral and \emph{vice versa}. 

The correspondence to QM will be first introduced, as we learned from Newton, for the ideal case of isolated particles  \cite{Dolce:2009ce} (free particles imply persistent periodicities), and then generalized to interacting particles  \cite{Dolce:tune,Dolce:AdSCFT} (\ie local modulations of periodicities). The correspondence with second quantization will be given in Sec.(\ref{second:quant}). For the sake of simplicity we will mainly consider (spinless) scalar particles.

\subsection{Derivation of the canonical formulation of quantum mechanics}

\subsubsection{ Elucidative example: the Black Body radiation }\label{BlackBody}

To introduce the idea of quantization by means of the constraint of intrinsic periodicity we consider the Black Body radiation  \cite{Dolce:Dice,Dolce:2010ij,Dolce:2010zz,Dolce:2009ce}, also for historical reasons. The idea is represented in Fig.(\ref{corponero}), where we imagine each component of the electromagnetic radiation as the vibrating string of a grand piano.

 We know from Planck that to each elementary electromagnetic field component (mode) of periodicity $T $, \ie of fundamental angular frequency $\omega = 2 \pi / T$, is associated the quantized energy spectrum $E_n = n \hbar \omega = n h / T $.  
In ECT, each component of the --- classical --- electromagnetic field must be ``overdetermined'' by imposing its fundamental periodicity $T$ as constraint. The Planck energy spectrum is therefore interpreted as the harmonic spectrum of a massless periodic phenomenon (vibrating string) of  fundamental time periodicity $T = h /E$. That is, according to the postulate of ECT, such a component of the electromagnetic field must be described as a intrinsic periodic phenomenon of temporal periodicity $T$, \ie as a vibrating string of fundamental period $T$. 

\begin{figure}\label{corponero}
	\centerline{ \includegraphics[width=16cm]{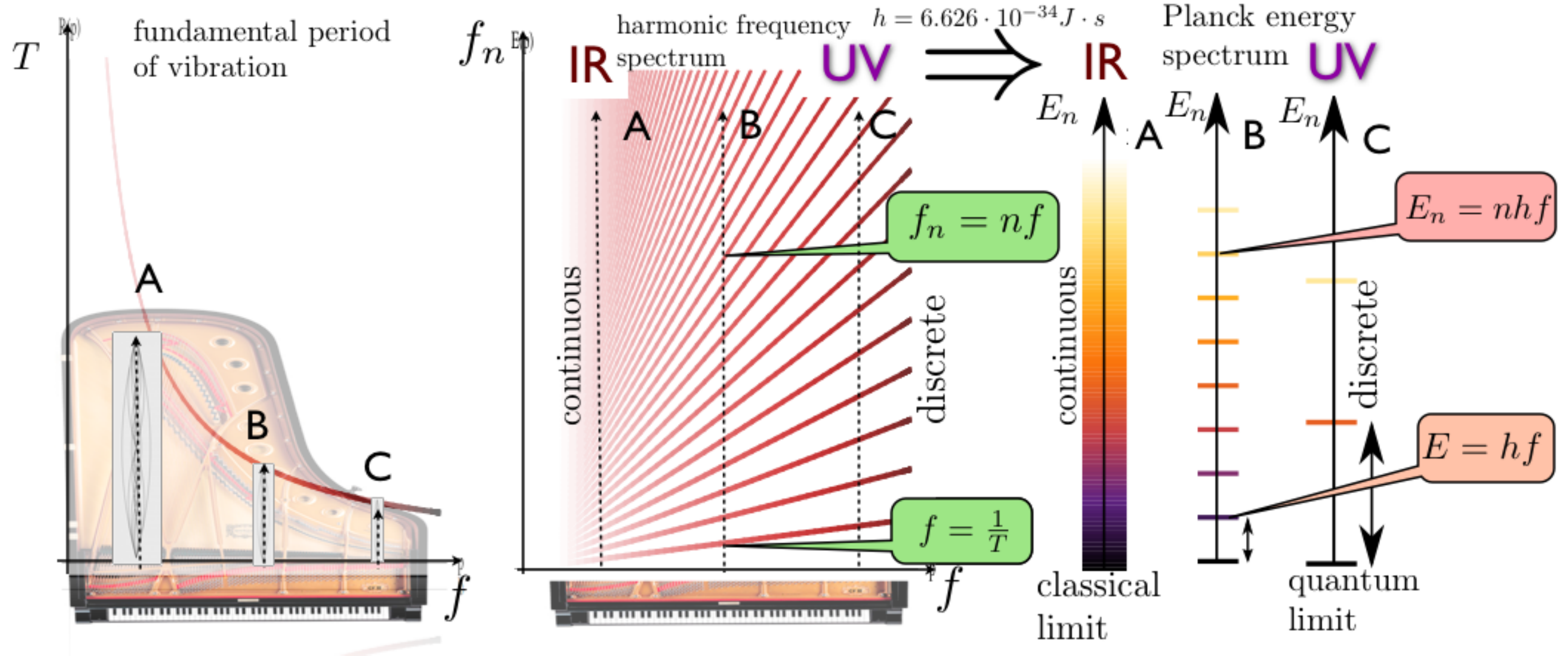}}
	%\includepdf[width=5cm]{harmonics.pdf}
	\caption{\footnotesize The electromagnetic field components A, B and C  can be imagined, according to ECT, as the strings of a grand piano vibrating with corresponding fundamental periods $T = 1 / f$ where $f$ is the generic frequency  of each field component. To each string is associated the frequency harmonic spectrum $f_n = n f$ which corresponds to the Planck energy spectrum $E_n = n h f = n h / T$, by means of the Planck constant. In our analogy to the low frequency notes (see letter A in the figure) corresponds infra-red (IR) components of the electromagnetic radiation of the Black Body. In this region the spectrum can be actually approximated to a continuum. The ultra-violet (UV) components, 
		corresponding to the high frequency notes (see letter C in the figure) of the grand piano, describes the portion of the spectrum which is manifestly quantized. These two limits describes  the classical limit and the quantum limit of the Black Body radiation, respectively.} \label{corponero}
\end{figure}

Indeed, such a vibrating string  has discrete frequency spectrum $f_n = n f = n/T$, Fig.(\ref{corponero}), which in turns, by means of the  Planck constant, yields  the ordinary Planck spectrum 
\begin{equation}E_n = h f_n  = n \frac{h}{T} = n E\,.  \end{equation}
In ETC the Planck quantization is described as the ``quantization'' (discretization) of the harmonics frequency spectrum of the string constrained to vibrate (\ie PBCs)  with corresponding fundamental period $T = h / E$. 

Similarly the constraint of spatial periodicity $\lambda^i = h / p_i$ of the electromagnetic field component (the field component wave-length) implies the quantized harmonic spectrum for the momentum 
\begin{equation}
p_{(n) i} = n p_i  = n \frac{h} {\lambda^i}\,. 
\end{equation} 

  Being a massless system (\ie infinite Compton periodicity), the temporal and spatial components of the four-periodicity are proportional,  $T = |\vec \lambda| /c$,  see (\ref{disp:relat}). In agreement with relativity we get the massless dispersion relation $E = |\vec p| c$ and the related massless dispersion relation of the energy spectrum 
  \begin{equation}
  E_n = \frac{n h} {T} = \frac{n h c}{ |\vec \lambda|} = c |\vec p_n|\,.
  \end{equation} 
  Notice that both the energy and momentum spectrum are labeled by the same quantum number $n$ as they are referred to the harmonics of the same ``periodic phenomenon'' (topology $\mathbb S^1$). 
  
  By ``overdetermining'' the massless relativistic dynamics with the constraint of intrinsic periodicity we have obtained the ordinary Planck's description of the Black Body radiation which, as well-known, avoids the ultra-violet catastrophe, see caption of Fig.(\ref{corponero}). 

To complete the description of the Black-Body radiation we will describe the role of the temperature in ECT in sec.(\ref{temperature}).

\subsubsection{Derivation of the axiom of the states}\label{states}

According to the postulate of intrinsic periodicity, in ECT every elementary (quantum) particle, \ie the basic constituents of nature,  is represented as a ``periodic phenomenon''/vibrating string/elementary clock. Let us now generically denote it as a function $\Phi(\vec x,t)$  of the cyclic space-time coordinates; or by $\Phi(x^\mu)$ in relativistic notation; or by $\Phi(x)$, suppressing the Lorentz index. %  Moreover, such a free elementary particle of persistent four-momentum $p_\mu$.  

The free elementary particle of persistent four-momentum $p_\mu$, if observed from an inertial reference frame, is postulated to have persistent periodicity $T^\mu$, according to the phase harmony relation $c p_\mu T^\mu = h$. The space-time periodicity $T^\mu$ is determined by Compton periodicity $T_C$ through Lorentz transformations, as much as $p_\mu$ is determined by the mass.

Intrinsic space-time periodicity means that the  function $\Phi(x^\mu)$ representing an elementary particle must satisfy the following constraint of periodicity, \ie PBCs, 
\begin{equation}\label{PBCs}
\Phi(x^\mu) = \Phi(x^\mu + T^\mu c)\,. \end{equation}

Let us consider for a moment only the time component of the periodicity. By means of \textit{discrete Fourier transform}, \ie in analogy with the vibrating string of periodicity $T$, the free relativistic particle is therefore represented as an harmonic system
\begin{equation}
\Phi(x) = \sum_n c_n e^{-\frac{i}{\hbar} E_n t }\phi_n(\vec x)\,, \label{el:part}   
\end{equation}   
$c_n$ are Fourier coefficients, $E_n = n h / T$ is the energy spectrum associated to the harmonic frequency spectrum $f_n = n / T$ of fundamental time periodicity $T$.
 We have a sum of all its possible harmonics, see previous papers and forthcoming one  \cite{Dolce:FPIpaths}. 
 
 Similarly, by considering the spatial component  $\lambda^i$ of the persistent space-time periodicity $T^\mu$, the free relativistic particle has harmonic momentum spectrum $p_{(n) i} = n h / \lambda^i$. Respectively, the temporal and spatial harmonics are  
\begin{equation}
\phi_n(t) =e^{-\frac{i}{\hbar} E_n t }\,~~~,~~~~~~\phi_n(\vec x) = e^{\frac{i}{\hbar} \vec p_n \cdot \vec x}\,. 
\end{equation} 

These are actually the quantized spectra prescribed by ordinary QFT (after normal ordering) for the free relativistic particle. Indeed the dispersion relations (\ref{disp:relat}) implies that the energy and the temporal period for a free particle varies  with the momentum as $E^2(\vec p) = \frac{h^2}{T^2 (\vec p)} = {\vec p^2 c^2 + M^2 c^4}$ so that the dispersion relation of the energy spectrum is\footnote{\label{foot} As a direct consequence of second quantization the (normally ordered) energy spectrum of an ordinary field mode of angular frequency $\omega(\vec p)$ is $E_n(\vec p) = n \hbar \omega(\vec p)$ with $n \in \mathbb N$. If we consider that a field has modes with positive and negative energy branches $\omega(\vec p) = \pm \sqrt{\vec p^2 c^2 + M^2 c^4}$ we immediately find the energy spectrum prescribed by ECT (\ref{disp:rel:spectr}). The negatives modes describes antiparticles which, for neutral bosonic particles, are indistinguishable from ordinary matter particles, whereas  for fermionic particles describes holes in the Dirac sea in agreement with ECT \cite{Dolce:Dice2014,Dolce:tune,Dolce:AdSCFT,Dolce:cycles}. } 
\beq\label{disp:rel:spectr}
E_n(\vec p) = n \hbar \omega_n(\vec p) = n \frac{h}{T(\vec p)} = n \sqrt{\vec p^2 c^2 + M^2 c^4}\,.
\eeq 

Notice again that both the energy and momentum spectra are described by the same ``quantum'' number $n$, with $n \in \mathbb Z$ --- we can assume  $n \in \mathbb N$ in most of the cases investigated in this paper (neutral scalar particles) see footnote(\ref{foot}). The temporal and spatial components of the space-time periodicity $T^\mu$ are not independent: they are the Lorentz projections (\ref{Lorentz:Tc}) of the single fundamental periodicity, \ie of the Compton periodicity (topology $\mathbb S^1$). These spectra are harmonic as $T^\mu$ is (globally) persistent  for free particles. 

\begin{figure}
\centerline{ \includegraphics[width=12cm]{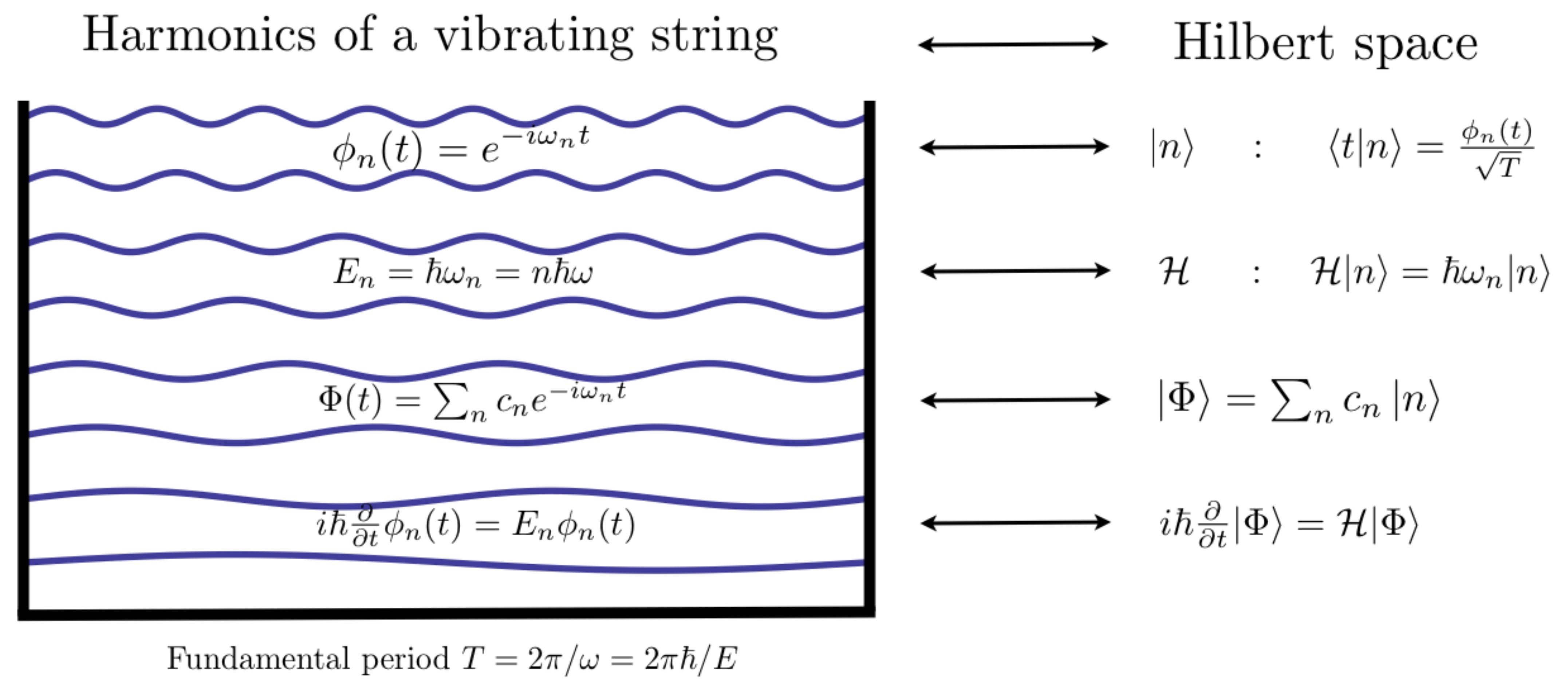}}
%\includepdf[width=5cm]{harmonics.pdf}
\caption{Harmonic vibrational modes $\phi_n$ of a string of fundamental period $T$. The state of the \textit{vibrating string} can be represented as a point  $|\Phi \rangle$ in a Hilbert space of basis $|n \rangle$.  The decomposition of the vibrating string into its vibrations in distinct \textit{harmonics} is given by the projection of the point onto the coordinate axes in the Hilbert space. The harmonic frequency spectrum $f_n = n f = n / T$  is represented by a corresponding "frequency" operator $ \mathcal H_f |n \rangle \equiv f_n |n \rangle$ in the Hilbert space. Similarly it is possible to define, at a classical level, by means of the Planck constant, the Hamiltonian operator $ \mathcal H$. The evolution law of the vibrating string in the Hilbert notation is given by the the Schr\"odinger equation.} \label{harmonics} %\label{ra_figo}
\end{figure}

From a historical point of view the mathematical concept of Hilbert space was originally introduced to describe classical harmonic systems, as described in Fig.(\ref{harmonics}) (see also wikipedia page ``Hilbert space''). It can be actually used to describe pure classical systems, contrarily to a common myth. 

Indeed, the harmonic modes $\phi_n(x)$ form a complete and orthogonal set which can be used to define the basis of the Hilbert space $\mathsf H$ associated to the elementary particle $\Phi$. Thus to every harmonic $\phi_n(x)$ of the elementary particle is associated the basis vector $| n \rangle$ of the Hilbert space 
\begin{equation}
\phi_n(\vec x) ~~~~~~~ \rightarrow ~~~~~~~  | n \rangle ~~~~ (~\text{or} ~~ |\phi_n \rangle  ~~  )\,,
\end{equation}
defined as
\begin{equation} \label{eigen:period}
\langle \vec x |n \rangle = \frac{\phi_n(\vec x)}{\sqrt{V_\lambda}}\,, 
\end{equation}
where $V_{\vec \lambda}$ is the volume spanned by a single spatial period (wave-length)  $\vec \lambda = \vec T c$.

The completeness relation of the harmonics $\phi_n(x)$ defines the inner product of the corresponding Hilbert space $\mathsf H$ 
\begin{equation} \label{ortog:period}
\langle n | n' \rangle = \int_{V_{\vec \lambda}} \frac{d^3 x}{V_{\vec \lambda}} \phi^*_{n'} (\vec x) \phi_n (\vec x) = \delta_{n,n'}\,. 
\end{equation}

The basis vectors $| n \rangle$  and the inner product $\langle \,\,\, | \,\,\, \rangle $  define the Hilbert space associated to the free elementary particle.

We may equivalently consider as integration volume that associated to an arbitrary large integer number of periods $N \vec \lambda$ (with $N \in \mathbb N$). The corresponding normalization is given by substituting the volume $V_{\vec \lambda} \rightarrow V_{N \vec \lambda}$ in (\ref{eigen:period}, \ref{ortog:period}).  It is however convenient to normalize over an infinite number of periods,  \ie over an infinite integration volume, as in ordinary QFT. In this case the substitution in the normalization factor must be  $V_{\vec \lambda} \rightarrow 2 \pi$:  
\begin{equation} \langle n | n' \rangle = \int \frac{d^3 x}{2 \pi} \phi^*_{n'} (x) \phi_n (x) = \delta_{n,n'}\,. \end{equation}
and 
\begin{equation} \langle x | n \rangle = \frac{\phi_n(x)}{\sqrt {2 \pi}}\,. \end{equation}
\emph{Thus the elementary space-time cycle associated to the elementary free particle naturally defines  a corresponding Hilbert space $\mathsf H$}. 

At this point it is straightforward to see that the elementary particle is represented by a vector in the corresponding Hilbert  space, \ie by a Hilbert state $|\Phi \rangle$. 
In fact, $\Phi(x)$, see (\ref{el:part}), is the superposition of all the possible harmonic modes allowed by the condition of intrinsic periodicity. Therefore, in the Hilbert space $\mathsf H$, \emph{the elementary space-time cycle is described by the corresponding Hilbert state} $|\Phi \rangle$:
\begin{equation}\Phi(x) ~~~~~~~ \rightarrow ~~~~~~~   |\Phi \rangle = \sum_n c_n  |n \rangle\,.  \end{equation}

The Hilbert description can be easily generalized to interacting particles, see Sec.(\ref{interaction}). More in general, every isolated system $\mathsf{S}$ in nature can be decomposed in terms of a finite number $N_{pt}$ of elementary particles $\Phi_i$ with $i = 1 ,2 \dots, N_{pt}$ (and their reciprocal interactions). Every elementary particle $\Phi_i$ of the system can be represented as a state in the corresponding Hilbert space $\mathsf H_i$. This means that the system is represented by the state $|\Phi_{\mathsf{S}} \rangle = |\Phi_1, \Phi_2, \dots, \Phi_{N_{pt}} \rangle$ defined in the Hilbert space resulting from the tensor product of all the Hilbert spaces of the elementary particles $\mathsf H_{\mathsf{S}} = \mathsf H_{1} \otimes \mathsf H_{2} \otimes \dots \otimes \mathsf H_{N_{pt}} $. 

Indeed a harmonic system characterized by more independent fundamental periodicities is described by the tensor product of the corresponding Hilbert spaces (Fock space). This is a general property of classical harmonic systems.  For example, we  will also see that the atomic orbitals are described by the product of the locally deformed periodicity in space-time associated to the Compton clock of the electron with Hilbert basis $|n \rangle$ and the spherical periodicity (\ie the vibrations of a spheric membrane) with Hilbert basis $|l,m \rangle$. Thus the Hilbert basis of the atomic orbitals is given by the corresponding vectorial product $|n \rangle \otimes |l, m \rangle$. 

In this way we have inferred the axiom of the state of ordinary QM directly from the  postulate of intrinsic periodicity,: 

\textbf{Theorem (Ex Axiom I of QM):} \textit{ To every  system $\mathsf{S}$ is associated a Hilbert space $\mathsf H_{\mathsf{S}}$ and $\mathsf{S}$ is represented by the Hilbert state $| \Phi_{\mathsf{S}} \rangle$.}

\subsubsection{Derivation of the axiom of the observables}\label{observables}

Even considering classical mechanics, to the harmonic frequency spectrum $f_n = n f = n / T$ of the classical string vibrating with fundamental periodic $T$ can be associated a frequency operator $\mathcal H_f$ in the corresponding Hilbert space, see (\ref{el:part}) and Fig.(\ref{harmonics}).   It returns the corresponding frequency eigenvalue when applied to the basis element $|n \rangle$ (the n$^{th}$ harmonic) : $\mathcal H_f |n \rangle \equiv f_n |n \rangle$. This defines at a classical level the frequency operator $\mathcal H_f$. 

Similarly to the case of the ``quantized'' harmonic frequencies of a vibrating string, to the n$^{th}$ harmonic $\phi_n$ of the elementary vibrating string (free elementary relativistic particle) is associated a quantized energy $E_n = h f_n =  n h / T $ and momentum $p_{(n) i} = n h / \lambda^i$. In the Hilbert space notation this defines the Hamiltonian and momentum operators $\mathcal H$ and $\vec \mathcal P$, respectively. They return the corresponding energy or momentum eigenvalues when applied to the corresponding harmonics: 
\begin{equation} \mathcal H | n \rangle  \equiv E_n |n \rangle ~, ~~~~~~~ \vec \mathcal P | n \rangle  \equiv \vec p_n | n \rangle\,. \end{equation}

In other words in ECT the Hamiltonian and momentum operators encode, at a classical level, the quantized spectra of the elementary space-time vibrations associated to the elementary particle.  

These are the fundamental operators of QM. The other operators can be constructed from them (all the physical quantities are functions of the phase-space variables). It is easy to show that the condition of periodicity (PBCs) guarantees that these Hilbert operators are Hermitians (sef-adjoint). Thus, by generalizing to the system of particles $\mathsf{S}$, we have proven the axiom of the observables:

\textbf{Theorem (Ex Axiom IIa of QM):} \textit{ For every physical observable of  the system $\mathsf{S}$ corresponds a linear self adjoint operator on the Hilbert space  $\mathcal A$.}

Similarly, by generalizing the fact that the only possible vibrational modes allowed by the fundamental periodicity to the string corresponds to  the ``quantized'' frequencies $f_n = n f$ we find the other axiom of QM:
  
\textbf{Theorem (Ex Axiom IIb of QM):} \textit{ All the possible results of the measurement of the observable $\mathcal A$ of  $\mathsf{S}$ are the eigenvalues $A_n$ of the operator.}

\subsubsection{Derivation of the axiom of the motion}\label{motion}

The Schr\"odinger equation is  the time evolution law of the vibrating string written in  the Hilbert space notation. It is straightforward to check that every harmonic --- which is actually a wave --- associated to the isolated periodic phenomenon, see (\ref{el:part}), satisfies the following differential equation (\ie the first order equation, ``square root'' of the wave equation)
\begin{equation}i \hbar \frac{\partial}{\partial t} \phi_n(t) = E_n \phi_n(t)\,. \label{wave:eq} \end{equation}

In ECT, the elementary particle (\ie the periodic phenomenon) is not however described by a simple wave. In general, as already said, it is the superposition of all the possible harmonics  allowed by its fundamental space-time periodicity. The evolution law of the ``periodic phenomenon'' with all its harmonics  is described by the Hilbert analogous of the time evolution equation (\ref{wave:eq}). In particular, it is easy to check that the time evolution law the persistent ``periodic phenomenon'', representing the free elementary particle, is actually given at a classical level by the ordinary time independent Schr\"odinger equation of QM
\begin{equation}
i \hbar \frac{\partial}{\partial t} |\Phi \rangle = \mathcal H |\Phi \rangle \end{equation}
The spatial evolution is derived in terms of the momentum operator in a similar way: $- i \hbar \frac{\partial}{\partial x^i} |\Phi \rangle = \mathcal P_i |\Phi \rangle$. As we will see, the time dependent Schr\"odinger equation refers to the interacting case (local modulations of periodicity). Thus we have derived the axiom of the motion:

\textbf{Theorem (Ex Axiom III of QM):} \textit{The time evolution of a quantum system $\mathsf{S}$, denoted by $\Phi_{\mathsf{S}}$, with  Hamiltonian operator $\mathcal H_{\mathsf{S}}$ is described by the differential equation on the corresponding Hilbert space $\mathsf H_{\mathsf{S}}$ :}
\begin{equation}i \hbar \frac{\partial}{\partial t} |\Phi_{\mathsf{S}} \rangle = \mathcal H_{\mathsf{S}} |\Phi_{\mathsf{S} } \rangle\,. \end{equation}

\subsubsection{Derivation of the axiom of the measurement}

Now we give a more accurate physical meaning of the ontological entity describing the elementary particle, that we have addressed as intrinsic ``periodic phenomenon'', internal clock or vibrating string, and is represented by $\Phi(x)$. We will find that in ECT, QM emerges as the statistical description of the ultra-fast, cyclic relativistic dynamics associated to the elementary particles due to the relatively poor temporal resolution  of modern timekeepers. 

The time periodicity $T$ of an elementary particle viewed in a generic inertial reference frame is always faster that the Compton periodicity $T \leq T_C$. This follows from the fact that the time period is determined by the energy $T = h / E$ whereas the Compton (rest) period is determined by the rest energy (mass) $T_C = h / M c^2$, and in relativity $E \geq M c^2$.  Furthermore, the heavier the mass of the particle, the faster the related Compton periodicity.  

For instance, even considering simple quantum-electrodynamics (QED), \ie systems of electrons interacting electromagnetically, the upper limit of the periodicities involved are determined by the Compton time of the electron --- the lightest massive particle of the Standard Model (expect neutrinos in which the ``oscillations'' associated to their masses are so slow that they have been indirectly observed, see neutrino oscillations). The Compton time of an electron is extremely small $T_{C_e} \sim 10^{-21} $s. This is the time taken by light to travel across the electron Compton length  $\Lambda_C \sim 10^{-12}$ m: that is, $ T_C =  \lambda_C / c $.

For comparison, the best resolution of modern timekeepers is only $\Delta T_{exp} \sim 10^{-17}$s, whereas the time scale of the Cesium atomic clock is $\Delta T_{Cs} \sim 10^{-10}$ s --- not to mention the temporal resolution in the Planck and de Broglie epoch. With such a poor temporal resolution  the direct observation of the ultra-fast cyclic dynamics of elementary particle is impossible. 

 Hence, only a statistical description of these ultra-fast periodic phenomena is possible\footnote{The difference of magnitude between the atomic clocks and the electron internal clock $ \Delta T_{C_{Cs}} / T_{C_{electron}} \sim 10^{11}$ is of the order of that between a solar year and the age of the universe$ \Delta T_{Universe} / T_{Seasons} \sim 10^{-11}$. Thus, trying to predict the outcomes of a quantum system is like trying to predict some annual dynamics having data with uncertainty in time of the order of the age of the universe. Due to the poor resolution in time the only possible way to describes these annual dynamics would be statistical.}.
 
 \emph{This is the key point for the derivation of the axiom of the measurement, and thus for the physical interpretation of the whole QM.  The mathematics of QM must be interpreted as the statistical theory describing at effective, low temporal resolution level the Compton ultra-fast cyclic dynamics of elementary particles. No hidden variables are involved, but only ultra-fast time dynamics. The Bell theorem or similar no go theorems based on hidden variables are not arguments applicable to rule out ECT.}
 
 This is similar to the observation of a rolling die without a slow motion camera: the outcomes can only be described statistically. ECT proofs that  the statistical description of the ultra-fast cyclic dynamics associated to elementary particles formally corresponds --- is formally equivalent ---  to QM.  The representation of elementary particles in the Hilbert space notation actually corresponds to give a statistical description of their cyclic dynamics, \ie to give up with a ``deterministic'' description due to the limits of the experimental techniques.  

In order to derive the Born rule from the postulate of elementary cycles it is necessary to consider that the description of very-fast intrinsic periodicity for an elementary particle is similar to the description of the dynamics of a particle moving very fast on a circle (elementary space-time cycles). Similarly to the ECT, it has been proven by 't Hooft that there is a close relationship between a particle moving on a circle and the temporal evolution of the harmonic quantum oscillator of the same period --- the time evolution of a ``periodic phenomenon'' This can be intuitively understood by thinking to a rolling die. A die is essentially a  ``periodic phenomenon'' on a lattice of $6$ sites ( it can be regarded also as a 't Hooft cellular automata). If the die rolls very fast and our resolution in time is too poor (\eg through a stroboscopic light, see Elze's stroboscopic quantization  \cite{Elze:2003tb}) we can only give a statistical descriptions of its outcomes.

According to ECT we can say that, as firmly believed by Einstein, ``God doesn't play dice''. In fact an observer with potentially infinite temporal resolution  would be actually able to resolve the ultra-fast cyclic dynamics associated to elementary particles, without invoking the statistical description (Hilbert notation) represented by QM.

Let us now consider  a similar example  \cite{Rosenstein:1985wb}. In a classical electric current we have electrons moving very slowly in a circuit. They can be thought of like  particles on a circle. The period $T$ now is large compared with our resolution in time. Due to the large number of electrons in the conductor, however, we are typically forced to give again a statistical description of the current by introducing a wave-function $\Phi(\vec x, t)$ satisfying the continuity equation \begin{equation}\label{eq:cont} \frac{\partial \rho}{\partial t} + \vec \nabla \cdot \vec j  = 0\,. \end{equation} 
We can assume that the charge density, \ie the density of electrons,  is given by $\rho = |\Phi|^2$; whereas $j$ is the current density. 

Of course, we can generalize this description to neutral classical (point-like) particles. In this case $\rho$  denotes the density of particles. It describes the probability to find particles in a given section of the circuit. Furthermore, we can generalize this statistical description even to the case of \emph{the single} particle in the circuit/circle, provided that it moves too fast  with respect to our timekeeper (\ie too small period $T$). In this case, similarly to a rolling die, we cannot determine its motion in detail (``deterministically'').  Again, we can only give a statistical description. 
Hence, the ``periodic phenomenon'' associated to \textit{the single} elementary particle in ETC is described, at a statistical level, in analogy to the current of a single classical particle moving very fast on a circle (circuit). Indeed, in ECT $\Phi(x)$  is described by a wave equation so that it satisfies the continuity equation (\ref{eq:cont}) --- as the observation of a particle implies that it must be stopped on a detector here we can assume the non-relativistic continuity equation \cite{Rosenstein:1985wb}. 
By hypothesis,  as the particle has a cyclic motion with fixed periodicity (PBCs with unknown small period $T$),  it can be written as the superposition of harmonic eigenstates, through discrete Fourier transformation, along with all the axioms derived so far.  

Finally we also have an important condition for $\Phi(x)$, \ie for the periodic phenomenon associated to the elementary particle: as we have \emph{a single particle} on the circle (circuit), the integral of the particle density $\rho = |\Phi|^2$ over the whole circuit must be equal to one, or, by integrating over an infinite number of period (\ie by using the $2 \pi$ normalization of the wave-function), we have the Born condition 
\beq
\int d x^3 |\Phi(\vec x, t)|^2 =  \sum_n |c^A_n|^2 = 1\,.
\eeq
  This describes the Born rule for our ``periodic phenomenon'' associated to the elementary particle.  

Thus we also have derived the axiom of the measurement from the postulate of intrinsic periodicity: 

\textbf{Theorem (Ex Axiom IVa of QM):} \textit{ For a  system $\mathsf{S}$ represented in the Hilbert space $\mathsf H_{\mathsf{S}}$ by $|\Phi_{\mathsf{S}} \rangle = \sum_n c^A_n |\phi^A_n\rangle $, where $|\phi^A_n\rangle$ are eigenstates of an  observable $\mathcal A$ (\ie eigenvectors of $\mathcal H_{\mathsf{S}}$), the probability to measure the eigenvalue $A_n$ of the observable $\mathcal A$ is given by $\mathcal P[A_n] = |c^A_n|^2$, such that  $\sum_n |c^A_n|^2  \equiv 1$.  }

 Moreover, since we are describing essentially a classical system we must assume that ``immediately after'' the measurement our ``periodic phenomena'' are in the state in which they have been observed. 

\textbf{Theorem (Ex Axiom IVb of QM):} \textit{ For a system $\mathsf{S}$ represented in the Hilbert space $\mathsf H_{\mathsf{S}}$, if the result of the measure of the observable $\mathcal A$ was $A_n$, ``immediately after''  this measure the state is in $|\phi^A_n\rangle$}. 

As the canonical formulation of QM is defined by all these axioms and we have derived all of them from the postulate of intrinsic periodicity of elementary particles, \emph{we conclude that ECT is formally equivalent to ordinary QM}. No hidden variables have been involved but only fast cyclic dynamics. See below for the generalization to the interacting case, and further proofs. 

\subsubsection{Derivation of the  commutation relations: Dirac quantization rule}

To derive the ordinary commutation relations of QM from the postulate of intrinsic periodicity we evaluate the ``expectation value'' of the generic observable gradient $\vec \nabla \mathcal F(x)$ for a generic elementary periodic phenomenon, that is $\langle \Phi |  \vec \nabla \mathcal  F(x) |\Phi \rangle$ --- the demonstration is analogous to the one used by R. Feynman \cite{Feynman:1948art} to prove that the path integral formulation of QM is equivalent to the canonical formulation of QM. 

The expectation value corresponds to weight the given observable by the Fourier coefficients $c_n$ of the harmonic system describing the elementary particle (the statistical meaning of the expatiation value is given by the the Born rule obtained above). For convenience here we use the normalization over a single period but the result can be easily generalized to large or infinite number of periods. By means of integration by parts and by using the definition of the momentum operators  we find %In other words we want to evaluate following expression
\begin{eqnarray}\label{expt:value}\langle \Phi |  \vec \nabla \mathcal  F(x) |\Phi \rangle &=& \frac{i}{\hbar} \int_{V_{\vec \lambda}} \left \{  \frac{d^3 x}{V_{\vec \lambda}} \sum_{n,m}  c^*_m c_n  e^{ - \frac{i}{\hbar} \vec p_m \cdot \vec x} (\vec p_m  \mathcal F(x) -  \mathcal F(x)  \vec p_n )  e^{\frac{i}{\hbar} \vec p_n \cdot \vec x} \right . 
 \nn \\ 
&-& \left .  \frac{i}{\hbar V_{\vec \lambda}} [\Phi^*(x)\mathcal F(x) \Phi(x)]_{V_{\vec \lambda}} \right \} = \frac{i}{\hbar}  \langle \Phi |   \vec{\mathcal P} \mathcal F(x) -  \mathcal F(x)  \vec{ \mathcal P}|\Phi \rangle\,. \end{eqnarray}

In this derivation it is very important to notice the crucial role of the assumption of spatial periodicity: here $x^i$ is a periodic variable of period $\lambda^i$. This guarantees that the boundary term of (\ref{expt:value}), obtained from the integration by parts, vanishes. That is, it guarantees that for arbitrary states $\Phi$ we obtain 
\begin{equation} i \hbar \frac{\partial \mathcal F(x)}{\partial x^i } =  \mathcal F(x)  \mathcal P_i - \mathcal P_i \mathcal F(x) = [\mathcal F(x),  \mathcal P_i] \,.  \end{equation}
As can be easily seen by choosing $\vec{\mathcal F}(x) = \vec x_j$, this yields the ordinary commutation relations of QM   
\begin{equation} [x_j, \mathcal P_i] = i \hbar \delta_{i,j}\,. \end{equation}

 This remarkable result shows that the commutation relations  of ordinary QM (and thus the Heisenberg uncertainty relation, see below) are implicit in the postulate of intrinsic periodicity --- \textit{vice versa} we can say that in the Dirac quantization rule, the commutation relations imposed to quantize the system implicitly hide the condition of intrinsic periodicity (commutation relations are typical of angular variables). 

Similar commutations relations can be extended to the other observables and to generic Hilbert states of system of periodic phenomena. Thus we obtain  the Dirac quantization rule of the canonical QM:

\textbf{Theorem (Dirac quantization rule)}: \textit{ If the commutation relation of the physical observables $A$ and $B$ of a system of classical particles is described by the Poisson brackets $\{A, B \}_P $, then the related system $\mathcal S$ [system of periodic phenomena] is described by the commutation relations}
\begin{equation}
[\mathcal A, \mathcal B] = i \hbar  \{A, B \}_P \,,
\end{equation} 
\textit{where $\mathcal A$ and  $\mathcal B$ are the corresponding operators of the Hilbert space $\mathsf H_{\mathsf{S}}$}.

\subsubsection{ Heisenberg uncertainty relation }
Even though the \textit{Heisenberg uncertainty relation} is implicit in --- is a direct mathematical consequence of --- the commutation relations derived above, it can also  be heuristically inferred directly from the assumption of intrinsic periodicity. 

We have just to  consider that the phase of the ``periodic phenomenon'' is defined modulo factors $2 \pi $'s,  and that in ECT the space-time coordinates are treated has periodic (angular) variables of period $T^\mu$ \cite{Nielsen:2006vc}. For the sake of simplicity, we only consider  the fundamental harmonic of the periodic phenomenon spatial evolution, \ie the monochromatic wave $e^{-\frac{i}{\hbar} p_i   x^i} $ of spatial periodicity $ \lambda^i = T^i c = h / p_i$ in the $i$-th spatial direction, and the phase invariance $2 \pi n'$ with $n' = 1$ (these give the most stringent uncertainty relation). Thus, in the observation of such a  ``periodic phenomenon'', the phase invariance $2 \pi$ \emph{implies a simultaneous uncertainty in the determination of the momentum and the position during its cyclic evolution}:
\begin{equation}e^{-\frac{i}{\hbar} p_i   x^i} = e^{-\frac{i}{\hbar} (p_i  x^i +  2 \pi \hbar)} =  e^{-\frac{i}{\hbar} (p_i + \Delta p_i) x^i }  =  e^{-\frac{i}{\hbar} p^i ( x^i + \Delta x^i)}\,, \end{equation}
where the simultaneous uncertainties are $\Delta p_i =  h / x^i$ and $\Delta x^i =  h / p_i$.

According to the postulate of elementary cycles this evolution is characterized by an intrinsic periodicity: that is, $\lambda^i$ is defined modulo wave-lengths $\lambda^i$, so that  $0 \leq ( x^i  ~~~ \text{mod} + n \lambda^i) \leq \lambda^i$,
%$( x^i  ~~~ \text{modulo} ~ \lambda^i) \in (0, \lambda^i]$ , 
where $p_i \lambda^i = 2 \pi \hbar$ and $n \in \mathbb N$. Hence, such ambiguity  is governed by the ordinary Heisenberg relation of QM
\begin{equation}\Delta p_i  \Delta  x^i = \frac{(2 \pi \hbar)^2}{ p_i   x^i} \ge \frac{(2 \pi \hbar)^2}{ p_i   \lambda^i} = h\,. \end{equation}

In general, by considering the commutation relations obtained above directly from the postulate of intrinsic periodicity, and by following the same demonstration of ordinary QM,  we find that:

\textbf{Theorem (Heisenberg uncertainty relation)}: \emph{In quantum measurement of the momentum and position (or similar conjugate quantities) of a particle the standard deviation of position $\sigma_{x^i}$ and the standard deviation of momentum $\sigma_{p_i}$ are related by the inequality:} 
\begin{equation}
\sigma_{p_i}  \sigma_{x^i}  \ge   \frac{ \hbar}{2}\,. 
\end{equation}

\subsubsection{Full generalization to interacting quantum systems}\label{interaction}

To generalize the formal correspondence between ECT and  QM to the case of interactions it is sufficient to consider that in undulatory mechanics the energy-momentum and the space-time periodicity are two faces of the same coin  \cite{Dolce:tune}. The phase-harmony condition now is local, it must be satisfied  in each interaction space-time point $X$. 

 Interactions are local variations of the energy-momentum. If we denote by $p_\mu$ the four-momentum of a free classical particle, when interaction is switched on, the resulting local four-momentum in the point $X$ is modified to $p'_\mu(x)|_{x=X} = \Lambda_\mu^\nu(x)|_{x=X} p_\mu$. The function  $\Lambda_\mu^\nu(x)$ describes, point by point, how the four-momentum varies due to interaction with respect to the free case.  Contrarily to the free case in which the periodicities are persistent (Newton's first law), interactions correspond to local modulations of the space-time periods $T^\mu(X)$  such that in  $X$ the phase harmony (de Broglie-Planck relations) is locally satisfied  
 \begin{equation}
 \oint_{T^\mu(X)} p'_\mu(x) dx^\mu = 2 \pi \hbar\,.
 \end{equation}

In analogy with modulated signals, the time and spatial evolutions of the ``periodic phenomenon'' are thus described by the modulations of all the harmonics  allowed by the local PBCs (in analogy with modulated waves)
\begin{equation}\label{wave:modul}
\Phi(\vec x, t'')\! = \! \sum_n c_n e^{-\frac{i}{\hbar} \int_{t'}^{t''}\!\! E'_n(x) dt } \phi_n(\vec x, t')\,, \;\;\;\;~~  \phi_n(\vec x'', t)\! = \! e^{-\frac{i}{\hbar} \int_{x'}^{x''} \!\! \vec p'_n(y) \cdot d \vec y}  \phi_n(\vec x', t)\,.
\end{equation}

We mention that the local modulations of space-time periodicity resulting from interactions is technically obtained by locally deforming the cyclic space-time dimensions of the periodic phenomena, or by local rotations of the space-time boundary. Surprisingly this leads to a general geometrodynamical description of interactions similar to general relativity. In addition to gravity it describes in a unified way also gauge interactions, depending whether the metric is locally curved, or it remains flat and only the boundary is locally rotated, respectively (the latter is a peculiar property of ECT)  \cite{Dolce:tune}.

In analogy with the free case, the locally modulated harmonics $\phi_n$ form a complete, orthogonal local set. In every point they define the basis of a local Hilbert space with corresponding local inner product denoted by $\langle ~|~ \rangle_X $. The local energy and momentum spectra $E'_n(x)$   and $p'_n(x)$ define local Hamiltonian and momentum operators $\mathcal H'(x)$ and $\vec {\mathcal P}'(x)$, respectively. 

 We introduce  the (persistent, \ie global) four-momentum operator of the free case $ {\mathcal P}_\mu = \{ \mathcal H, - \vec {\mathcal P}\}$. Similarly, if interaction is assumed, we introduce the local four-momentum operator $ {\mathcal P(x)}_\mu = \{ \mathcal H(x), - \vec {\mathcal P}(x)\}$ in terms of the local Hamiltonian and momentum operators. The local four-momentum operator of the interacting system is deduced from the free case through the same transformation introduced for $p'_\mu(x)|_{x=X}$:  the local transformation is $\mathcal P'_\mu(x)|_{x=X} = \Lambda_\mu^\nu(x)|_{x=X} \mathcal P_\mu$  \cite{Dolce:tune}.  Clearly this coincides with the Hamiltonian and momentum operator of ordinary QM for that interacting system, see also (\ref{lagran:FPI}). 

 The axiom of motion can be easily generalized to interactions by using the analogy with modulated waves, see (\ref{wave:modul}). The time evolution of  modulated cycles is given by the ordinary Schr\"odinger equation of the interacting system of local Hamiltonian $\mathcal H'(X)$:
\begin{equation}i \hbar \frac{\partial}{\partial t} |\Phi \rangle = \mathcal H'(X)|\Phi \rangle \,.  \end{equation}

An interacting system is characterized in every point of its cyclic evolution by a locally modulated spatial  periodicity. Thus the local commutation relations of the interacting system periodic phenomena can be easily inferred by following the same demonstration of the free case  
\begin{equation} [x, \mathcal P'] = i \hbar \,. \end{equation}

\emph{In this way the exact correspondence of ECT to QM (axioms of QM and Dirac quantization rule) has been generalized to interactions.} 

In conclusion, all the axioms of the canonical formulation of QM, as well as the Dirac quantization rule (commutation relation and Heisenberg relation) have been derived directly from the postulate of intrinsic periodicity. With this, \emph{we have fully proven the exact  equivalence between ETC and QM}. As further (unnecessary) check of this result we will next derive the equivalence to the Feynman path integral. Then we will also prove the equivalence to second quantization. What else should we prove?

\subsection{Derivation of the Feynman formulation  of quantum mechanics}

Now we prove, independently from the previous correspondences, another amazing aspect of ECT \cite{Dolce:cycles,Dolce:ECunif,Dolce:FPIpaths,Dolce:TM2012,Dolce:Dice2012,Dolce:cycle,Dolce:ICHEP2012,Dolce:tune,Dolce:AdSCFT,Dolce:FQXi,Dolce:Dice,Dolce:2010ij,Dolce:2010zz,Dolce:2009ce,Dolce:EPJP,Dolce:SuperC,Dolce:Dice2014}: 

\textbf{Theorem (Feynman Path Integral)}: \emph{The classical-relativistic cyclic  dynamics of ECT directly implies the ordinary Feynman Path Integral and vice versa}. 

\begin{figure}
	\centerline{~~~~~~~ ~~~~~~~ ~~~~~~~ \includegraphics[width=12cm]{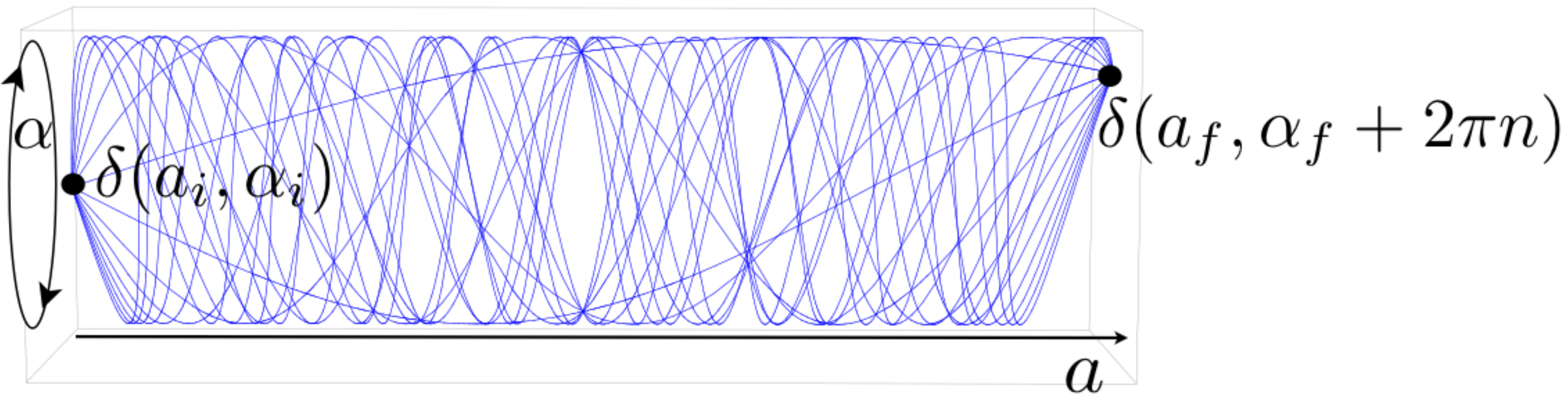}}
	\caption{A representation of periodic classical paths (blue lines)  between two extremal points $\delta_i$ and $\delta_f$ of a cyclic geometry of axial coordinate $a$ and cyclic coordinate $\alpha$. In general we have infinite set of degenerate periodic paths. Here we have only reported those with winding numbers $n'=-7, - 6, \dots , 6, 7$. In analogy with this picture, the evolution of a relativistic quantum particle is given by the interference of all the classical periodic paths on the corresponding cyclic space-time geometry of ETC. Remarkably the interference of periodic paths exactly described by the ordinary Feynman path integral.} \label{fig1}
\end{figure}

\begin{figure}
	\centerline{a)  \includegraphics[width=6cm]{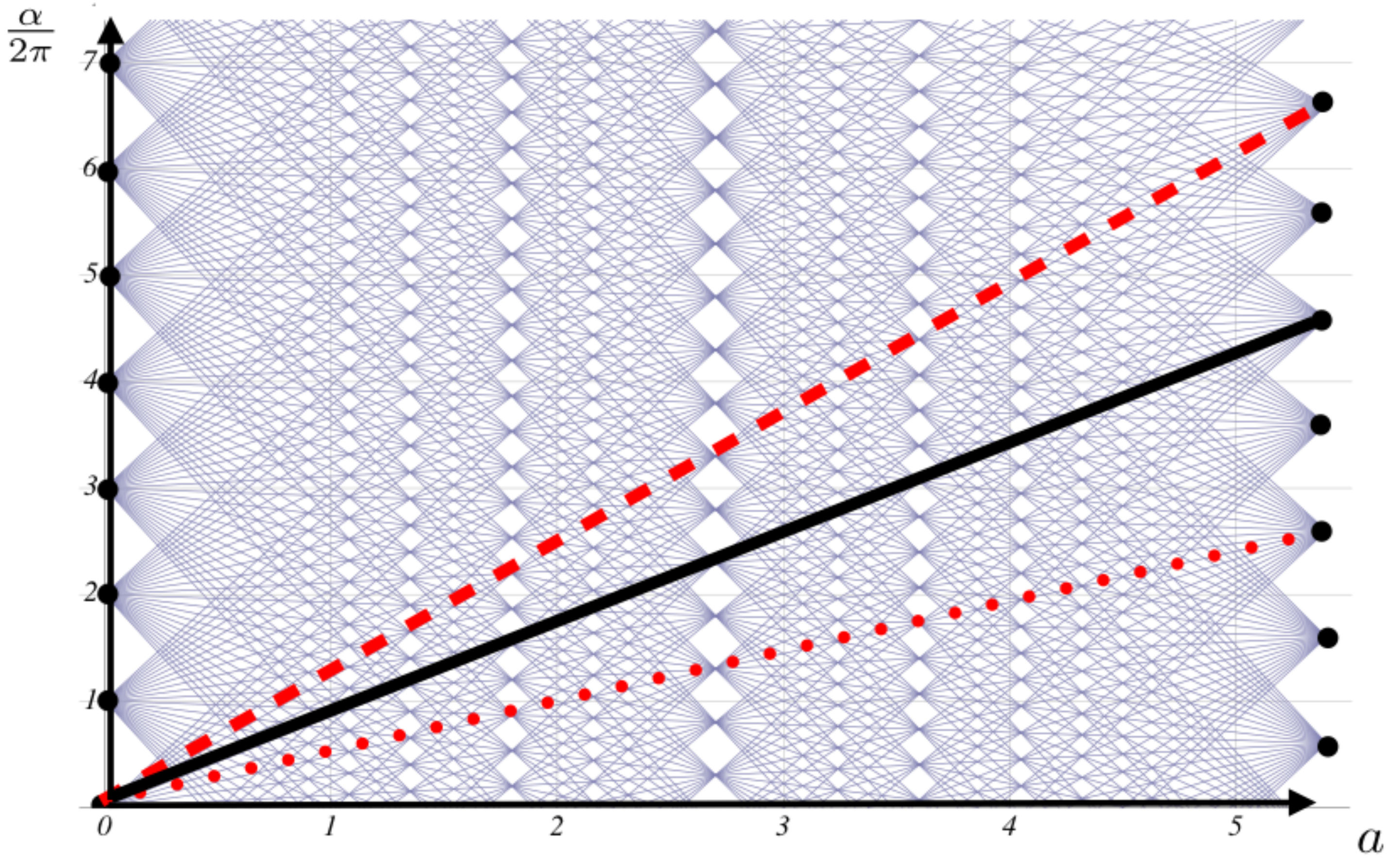} b)\includegraphics[width=6cm]{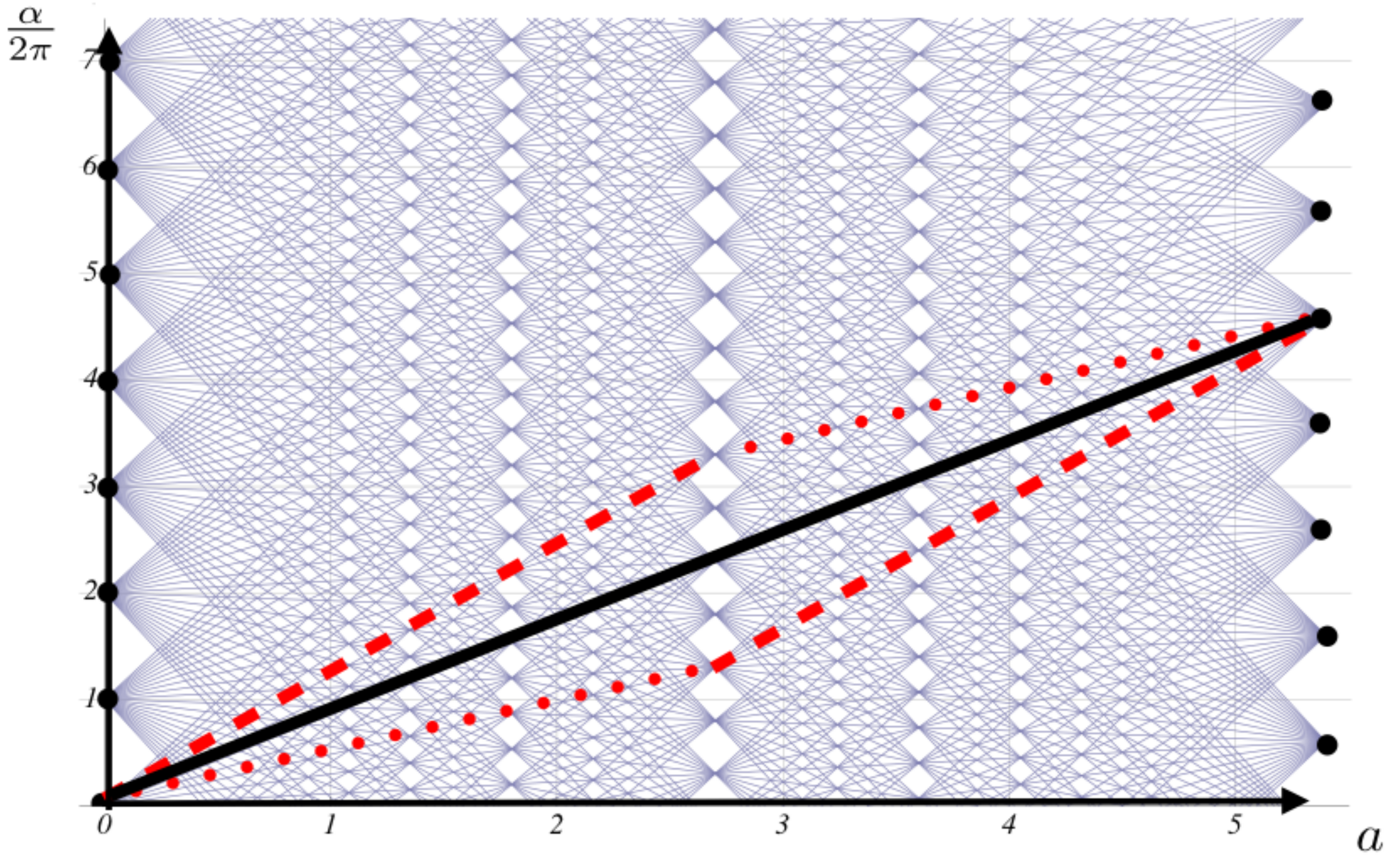}}
	\centerline{~~~~~~~~~~ c) \includegraphics[width=12cm]{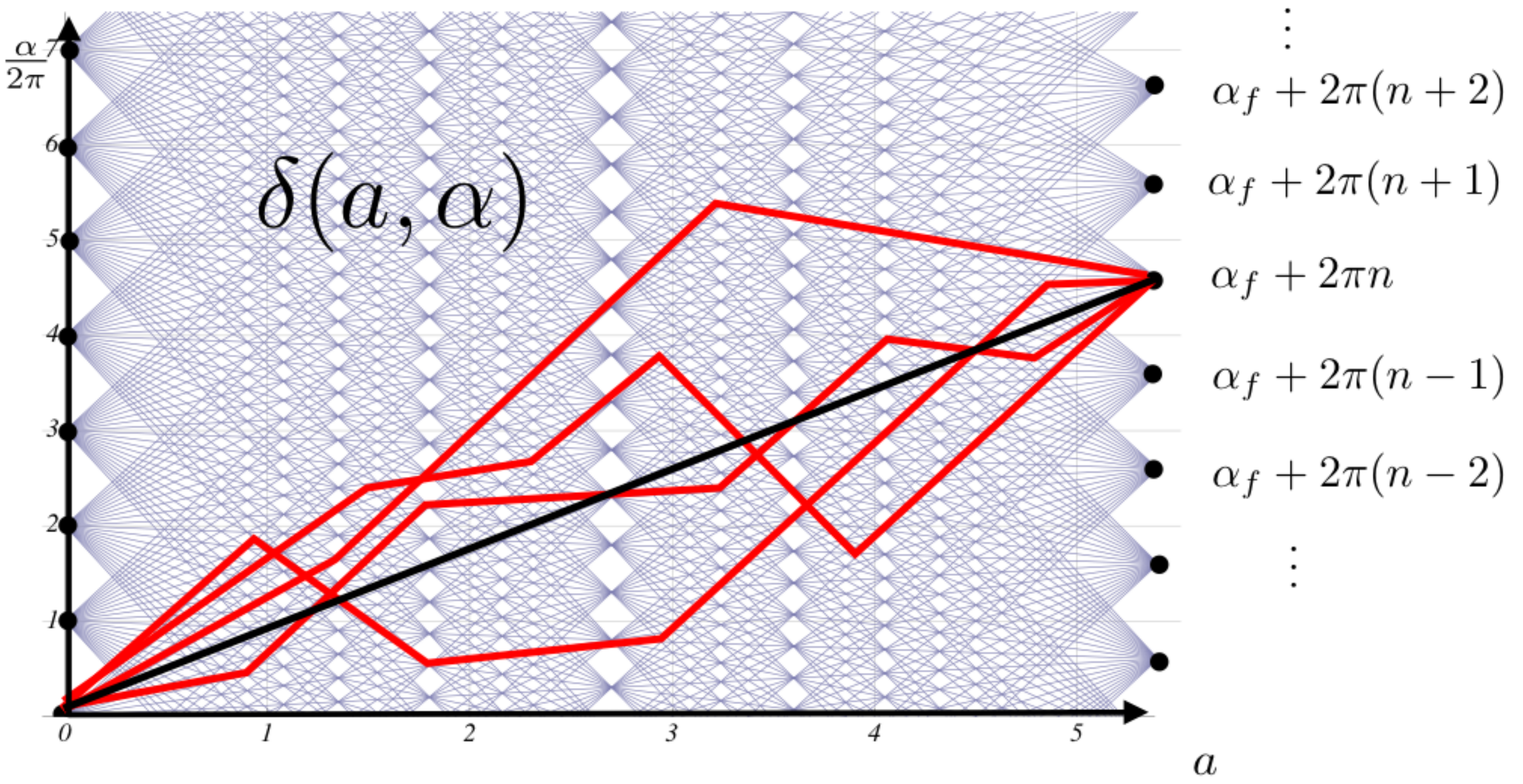}}
	\caption{We imagine to open the cylinder of Fig.\ref{fig1} on a Cartesian plane $(a, \alpha)$, where $\alpha$ has periodicity $1$. The periodic classical paths of the cyclic geometry are again represented by blue lines.  In Fig.(a) we arbitrarily choose two of these periodic paths, represented as dotted and dashed red lines. The black line represents the classical path of an ordinary particle evolution, on the ordinary non-compact space-time, between the extremal point $(a_i, \alpha_i)=(0,0)$ and  $(a_f, \alpha_f)$.  Due to the unitarity of the cyclic evolution each of these paths can be cut and translated by periods, as in Fig.(b) obtaining variations around the classical path b). Despite the cylindric geometry considered is a simplification of the cyclic space-time geometry, this shows that the degenerate paths of a cyclic evolution can be combined to form the Feynman variations relations around the ordinary classical particle path (c). %In general, \emph{the classical evolution of a space-time periodic phenomenon is described by the ordinary Feynman path integral}. By considered that the periodicity if fixed, we have that the interference is totally constructive only if the extremal points are the classical path, otherwise the interfere becomes more and more destructivee.
	} \label{fig2}
\end{figure}

Let us start with a digression. The evolution of ``periodic phenomena'' is similarly to the classical evolution between two points on a cylindric geometry,  Fig.(\ref{fig1}). It is characterized by an infinite number of degenerate classical paths with common extremal points. In Fig.(\ref{fig1}) we represent the degenerate evolutions on cylinder, rather than on a cyclic one, in order to elucidate the idea. The elementary cycle evolution is described by the interference  of all the possible paths with different winding numbers associated to the cyclic space-time geometry of ETC, see also Fig.(\ref{fig2}).

                                                              Clearly, this is a general property of vibrating strings. as it can be proven by using the \textit{Poisson summation} defined as
 \beq
 \sum_n c_n e^{-i n \omega (t'' - t')} = 2 \pi \sum_{n'} \hat c_n \delta ( \omega (t'' - t') + 2 \pi n') = T \sum_{n'} \hat c_n \delta ( (t'' - t') +  n' T)\,,
 \eeq
  where $n, n' \in \mathbb Z$ (see footnote.\ref{foot}), $\hat c_{n'}$ are the ``discrete'' Fourier transform coefficients of $c_n$. Clearly here $t$ is a periodic (angular) coordinate of period $T= 2 \pi /\omega$. In this example we have that a one dimensional periodic phenomenon, written as the sum of its harmonics, is equivalently represented by a sum of classical periodic paths of period $T$ and winding number $n'$. The periodic paths are described by the Dirac deltas and correspond to the degenerate paths between the initial point $t'$  and the degenerate final points $t'' +  n' T $.      


Similarly, by applying the Poisson summation to the persistent, relativistic periodic phenomenon, we find that the space-time evolution between $x'_\mu$ and $x''_\mu$ of the free relativistic particle is
\bea\label{wave:paths}
\Phi(x'',x') &=& \sum_n  c_n e^{-\frac{i}{\hbar} n [E (t'' - t') - \vec p \cdot (\vec x'' - \vec x')]  } \nn \\ &=&  2 \pi \hbar \sum_{n'} \hat c_{n'} \delta( E (t'' - t') -  \vec p \cdot (\vec x'' - x') + 2 \pi n' \hbar )\,.   
\eea 

Again, the sum over the Dirac deltas describes the interference among the classical-relativistic period paths labeled  by the winding numbers $n'$.   It describes the cyclic space-time evolution associated to the persistent periodic phenomenon. On the space-time cyclic geometry of ECT, we have an infinite set of degenerate classical space-time paths among two arbitrary extremal points $x'_\mu$ and $x''_\mu$. The analogous of these degenerate cyclic classical paths are represented by the blue lines in Fig.(\ref{fig1}) and Fig.(\ref{fig2}) --- the degeneracy of classical paths provides an alternative interpretation of the Heisenberg relation, because they have different energies, according to the quantization, and different arrival space-time points. 

To generalize this to interacting particles we have just to consider the generalization of the Poisson summation to curved space-time, in order to describe the locally deformed cyclic space-time geometry of elementary particles. 

Let us now formally derive the Feynman path integral from the postulate of intrinsic periodicity (without any farther assumption). We can directly consider the case of an interacting relativistic particle. We have seen that the space-time evolution of  the modulated ``periodic phenomenon'' is characterized in every point by a local Hilbert space with corresponding local inner product. The time evolution operator in the free case was $U(t'',t') = e^{-\frac{i}{\hbar} \mathcal H (t'' - t')}$.  According to Sec.({\ref{motion}}), in the interacting case the time evolution operator is the Hilbert analogous of the modulated wave 
\begin{equation}U(t'',t') = e^{-\frac{i}{\hbar} \int^{t''}_{t'} \mathcal H' (x) dt}\,.\end{equation}

The time evolution operator is unitary (and Markovian). Hence the time evolution can be sliced in $N \rightarrow \infty$ elementary time evolutions of infinitesimal duration $\epsilon$:
\beq
U(t'',t') = \lim_{N\rightarrow \infty} \prod_{m=0}^{N-1} U(t'+t_{m+1},t'+t_m - \epsilon)
\eeq
where the infinitesimal time evolution is
$$ U(t'+t_{m+1},t'+t_m - \epsilon) = e^{-\frac{i}{\hbar} \mathcal H' (t_m) \epsilon}$$

%\begin{verbatim}
%\begin{center}

For the sake of simplicity we assume the normalization over an infinite number of spatial periods --- it is sufficient that the integration region corresponds to a sufficiently high integer number of periods to completely contain the interaction region. In perfect analogy with Feynman's derivation,  we plug the local completeness relations of the modulated harmonic set $\phi_n$ of the modulated periodic phenomenon in between the elementary time evolutions.  By considering the local definitions of the Hamiltonian $\mathcal H'$ and momentum operator $\vec \mathcal{P'}$ in the Hilbert space we obtain
\begin{equation} \label{elem:path:sum}
U(x'',x') = \lim_{N\rightarrow \infty} \int \left(\prod_{m=0}^{N-1} {d^3 x_m}\right) U(x'',x_{N-1}) U(x_{N-1}, x_{N-2}) \dots U(x_1,x')\,, \end{equation}
where the infinitesimal space-time evolutions of the 	``periodic phenomenon'' are 
\begin{equation}U(x_{m+1},x_m) =  \langle \hat \Phi | e^{-\frac{i}{\hbar}(\mathcal H' \Delta t_m - \vec { \mathcal{P'}} \cdot \Delta \vec x_m  ) }|\hat \Phi \rangle 
\end{equation}  
with $|\hat \Phi \rangle = \sum_n | n \rangle$, $\Delta t_m = t_m - t_{m-1} $ and  $\Delta \vec x_m = \vec x_m - \vec x_{m-1} $. 
Without any further hypothesis than  intrinsic periodicity, we have derived the  ordinary Feynman path integral in phase-space
\begin{equation}
\mathcal Z = U(x'',x') = \lim_{N\rightarrow \infty} \left( \int \prod_{m=0}^{N-1} d^3 x_m\right) \prod_{m=0}^{N-1} \langle \hat \Phi | e^{-\frac{i}{\hbar}(\mathcal H'\Delta t_m - \vec {\mathcal P'} \cdot \Delta \vec x_m  ) }|\hat \Phi \rangle \,.
\end{equation}

The phase of the modulated ``periodic phenomenon'' defines the Lagrangian of the interaction 
\begin{equation}
\label{lagran:FPI} \mathcal L  = \vec {\mathcal P'} \cdot \vec {\dot x} - \mathcal H' \,. 
\end{equation}

By construction, according to the definition of the local Hamiltonian and momentum operators given in the previous paragraph, this Lagrangian is formally the Lagrangian of the interacting classical particle written in terms of operators. The action is 
\begin{equation}\mathcal S = \int_{t'}^{t''}\mathcal L dt = \int_{t'}^{t''} (\vec {\mathcal P'}  \cdot \vec {\dot  x} - \mathcal H') dt\,. \end{equation}
This yields the ordinary Feynman path integral in Lagrangian notation 
\begin{equation}\mathcal Z = \int \mathcal D^3 x e^{\frac{i}{\hbar} \mathcal S}\,. \end{equation} 
\emph{With this we have proven an exceptional result: the classical-relativistic  evolution of modulated elementary space-time cycles is equivalently described by the ordinary Feynman path integral of the corresponding interacting relativistic particles}. 

\emph{Vice versa}, the ordinary Feynman path integral explicitly  leads to and the intrinsically cyclic dynamics of ECT, as can be easily proven by means of  the Poisson summation. This also clearly shows the physical meaning of the correspondence between elementary cycles evolution and Feynman path integral. The idea is illustrate in Fig.(\ref{fig1}) and Fig.(\ref{fig2}) in analogy with a simple two dimensional cylindric geometry. We consider the free case, but the result can be as usual generalized to interactions by using the Poisson summation for curved space-time coordinates. 

Let us assume the ordinary FPI of QM for a relativistic particle (the result is the same if we consider a field, see for instance \cite{Dolce:FPIpaths}). According to second quantization a relativistic free particle has harmonic energy and momentum spectra:  $ \mathcal H |\phi_n \rangle = n E  |\phi_n \rangle $ and  $ \vec{\mathcal P} |n \rangle = n \vec p  |n \rangle $ ---  we can assume $n \in \mathbb Z$, see footnote.(\ref{foot}). Thus, similarly to (\ref{wave:paths}), the elementary space-time evolution of the ordinary Feynman path integral of QM can be written, by means of the Poisson summation, as sum of Dirac deltas (represented by the segments of paths in Fig.\ref{fig2})
\bea
U(x_{m+1},x_m) &=&  \langle \Phi | e^{-\frac{i}{\hbar}(\mathcal H \Delta t_m - \vec {\mathcal P} \cdot \Delta \vec x_m  ) }|\Phi \rangle = \sum_n  e^{-\frac{i}{\hbar} n (E \Delta t_m - \vec p \cdot \Delta \vec x_m )  } \nn \\
&=& 2 \pi \hbar \sum_{n'_m}  \delta ( E \Delta t_m - \vec p \cdot \Delta \vec x_m + 2 \pi n'_m \hbar)\,.   
\eea  

By substituting in (\ref{elem:path:sum}) and using the Dirac delta property $\int d^3 x_m \delta (\vec x_{m + 1} - \vec x_m)\delta(\vec x_m - \vec x_{m-1}) = \delta (\vec x_{m + 1} - \vec x_{m-1})$ we explicitly find, in agreement with (\ref{wave:paths}), that the Feynman path integral is given by the sum over all the possible periodic paths of persistent periodicity $T^\mu$:
\bea \label{Feyn:cyclicpaths}
\mathcal Z_{free} &=&   \lim_{N\rightarrow \infty} \left( \int \prod_{m=0}^{N-1} d^3 x_m\right) \prod_{m=0}^{N-1} 2 \pi \hbar \sum_{n'_m}  \delta ( E \Delta t_m - \vec  p \cdot  \Delta \vec x_m + 2 \pi n'_m \hbar) \nn \\
&=& 2 \pi \hbar \sum_{n'}  \delta( E (t'' - t') - \vec  p \cdot (\vec x'' - \vec x') + 2 \pi n' \hbar ) \,.  
\eea 

The demonstration, in this form (\ie based on infinitesimal space-time paths) can be generalized to interaction and fields. The Feynman path integral of an interacting system is interpretable as the sum (integral) over all the possible periodic paths with locally modulated periodicity $T^\mu(x)$ of the interaction scheme.  These are all the possible windings associated to the locally modulated cyclic geometry  --- local deformations of the elementary space-time cycles. 
 
 In a more advanced notation  \cite{Dolce:2009ce,Dolce:tune,Dolce:AdSCFT,Dolce:cycles} the elementary space-time cycles are defined as solutions of classical-relativistic field actions on compact space-time and PBCs. Due to the PBCs, the least action principle in ECT leads to an infinite set of degenerate solutions. Like in a game of mirrors the extremal points of the evolutions are in fact defined modulo space-time periods. In ordinary non-compact space-time we can only have a single path minimizing the relativistic action with fixed extremal points. On the contrary in ECT for each classical paths minimizing the action we have an infinite set of degenerate paths with the same property, see Fig.(\ref{fig1}). 
 
 These degenerates paths are exactly the periodic paths obtained in (\ref{Feyn:cyclicpaths}): they all are classical-relativistic paths and they are labeled by the  winding number $n'$. \emph{Thus the Feynman path integral can be equivalently described as the interference of classical periodic paths.  Due to the cyclic nature of particles dynamics the classical variational principle is fully preserved in QM}. 

This further remarkable result clearly proves that the quantum evolution of the relativistic particle, as described by the Feynman path integral, is given by the \emph{interference} of all the periodic classical paths associated to its local de Broglie-Planck (space-time) periodicity. 

Notice: once that the local space-time period $T^\mu(x)$ (the shape of our deformed cyclic space-time geometry) is assigned,  the corresponding  local (classical) energy-momentum $p_\mu(X)$, is assigned as well. In turn the classical particle path is assigned as well --- in this case we mean the ordinary particle classical path as obtained in the ordinary non-compact space-time, see  black line in Fig.(\ref{fig1},\ref{fig2}). The maximal probability is along the path of the corresponding classical particle, where all the periodic paths have constructive interference.   If the final or initial point $x''$ moves away from the classical path, the interference of the periodic paths contributing to the Feynman Path Integral becomes less and less constructive, denoting a lower probability to find the particle in that point,  in perfect agreement with Feynman's interpretation.

\subsection{Derivation of the Bohr-Sommerfeld quantization and WKB method}\label{Bohr-Sommerfeld}

Here we prove that the ordinary semi-classical formulations of QM are direct consequences of the postulate of intrinsic periodicity. The postulate of intrinsic periodicity is in fact the generalization, in terms of space-time geometrodynamics, of the wave-particles duality.

The constraint of persistent or modulated space-time periodicity is the quantization condition in ECT for the free or interacting case, respectively. In the free case persistent periodicity $T^\mu$, \ie the PBCs (\ref{PBCs}), implies
\begin{equation}
\Phi(x) = \sum_{n} c_n e^{-\frac{i}{\hbar} p_{(n) \mu} x^\mu} ~~\Rightarrow ~~ e^{-\frac{i}{\hbar} p_{(n) \mu} T^\mu c}= e^{-i 2 \pi n}  \,.
\end{equation}
The temporal and spatial components of this relation yields the (normally ordered) energy and momentum spectra for the free relativistic particle (similarly to a particle in a spatial and temporal box and in agreement with second quantization)
\bea\label{BS:pers}
e^{-\frac{i}{\hbar} E_{n} T}= e^{-i 2 \pi n} ~~~~\rightarrow ~~~&E_{n} T = 2 \pi n \hbar &  ~~~~\rightarrow ~~~ E_n = n \frac{h}{T}\,, \nn \\
e^{\frac{i}{\hbar} \vec p_{n} \cdot \vec \lambda}= e^{- i 2 \pi n} ~~~~\rightarrow ~~~&\vec p_{n} \cdot \vec \lambda  = 2 \pi n \hbar &~~~~\rightarrow ~~~ p^i_n = n \frac{h}{\lambda_i} \,.
\eea  

Interactions imply that in every  space-time point $X$ on which we are evaluating the particle evolutions is associated a locally modulated  space-time periodic $T^\mu(X)$, corresponding to the local momentum $p_\mu(X)$ of the interacting particle. In order to generalize the relation above we must consider that the ``periodic phenomenon'' associated to an interacting particle must be locally modulated, as in (\ref{wave:modul}). In this case the constraint of modulated periodicity (cyclic geometrodynamics) implies \cite{Dolce:tune}  
\begin{equation}\label{modul:solution}
\Phi(X) = \sum_n c_n e^{-\frac{i}{\hbar}\int^{X} p_{(n) \mu}(x) dx^\mu} ~~~\Rightarrow ~~~~ e^{-\frac{i}{\hbar} \oint_{T^\mu(X)} p_{n \mu} d x^\mu} = e^{-i 2 \pi n}  
\end{equation}
In the interaction point $x=X$ the deformed energy and momentum spectra are thus given by 
\begin{equation}\label{BS:energymom}
\oint_{T(X)} E_{n} d t = 2 \pi n \hbar ~~~~~~,~~~~~~~\oint_{\lambda(X)} \vec p_{n} \cdot d \vec x  = 2 \pi n \hbar \,.
\end{equation}  

This essentially means that the ``periodic phenomenon'' must have closed space-time orbits, in analogy, for instance, with Bohr orbitals. ETC is therefore a relativistic generalization of the \textit{Bohr-Sommerfeld quantization} and of the \emph{WKB method}. 

It must be noticed that the variational principle for theories in compact space-time generally allows for more general BCs. For instance, the relativistic bosonic action allows for  Dirichlet and Neumann BCs or twisted PBCs at the ends of  compact space-time dimensions, as well known in string and extra-dimensional theories. By assuming anti-periodicity in time, \ie $ \Phi(\vec x, t) = - \Phi(\vec x, t + T)$ (\eg Fermi-Dirac statistics), the resulting quantization of the energy is 
\begin{equation}e^{-\frac{i}{\hbar} E_{n} T}= - e^{-i 2 \pi n} ~~\rightarrow ~~E_{n} T = 2 \pi \left(n + \frac{1}{2}\right) \hbar   ~~\rightarrow ~~ E_n =\left(n + \frac{1}{2}\right) h f\,. \end{equation}
In this way we have also obtained the vacuum energy  $ E_0 =  h f / 2  $. %can be associated to antiperiodicity, avoiding the normal ordering assumed so far.  

 More in general the twist of an angle $2 \pi \alpha $ in the periodicity condition,  $\Phi(x) = e^{-i2 \pi \alpha} \Phi(x +Tc)$ implies the Morse term $\alpha$ in the Bohr-Sommerfeld quantization condition 
\begin{equation}\label{BS:morse}
\oint_{\lambda(X)}  p_{(n) \mu} d x^\mu  = 2 \pi (n + \alpha) \hbar \,.
\end{equation}

\emph{The vacuum energy} $E_0 = \alpha h f $ in the spectrum is the manifestation of a twist in the elementary space-time cycles constituting particles. The \emph{Morse term} or the vacuum energy are obtained  by analyzing the BCs of the system. For instance, the  zero-point energy of particles bounded in a potential it is determined by the BCs at the spatial infinite or from the spatial symmetry of the potential (\eg as in the harmonic potential or Coulomb potential).  In confirmation of this we must notice that, as also originally proposed by Casimir  \cite{Jaffe:2005vp}, the modern technique to calculate the \textit{Casimir effect}, manifestation of the vacuum energy, is actually based on the BCs of the system   \cite{Cugnon}.

 The semi-classical methods of QM are typically regarded as approximative description of QM. However ECT generalizes these semi-classical quantization methods giving them a full validity. We have in fact proven the equivalence of ECT with the main quantization methods. In simple words ECT reveals that the whole QM, including  the axioms of QM, the Feynman path integral, the Dirac quantization and, as we will see, second quantization has semi-classical origin. The postulate of intrinsic periodicity is the generalization of the wave-particles duality. Nevertheless ECT is radically different from the old quantum theory as the undulatory nature of quantum particles is directly encoded into the fabric of space-time geometry by means of compact space-time and related PBCs.

%As can be seen also by taking the spatial term of (\ref{Dirac:string}) 
%
%
% If we calculate, through the Stocke's theorem the magnetic flux in a circuit $\Sigma$ we find that the magnetic flux is quantized. This follows from the spatial components of the above equation
%\begin{equation}e^{\frac{i e}{\hbar} \oint_{\Sigma} A_{i} d x^i =  e^{-i 2 \pi n} ~~~~\rightarrow ~~~  \oint_{T(X)} A_\mu d x^\mu = 2 \pi n \hbar \,.\end{equation}
%
%
%the assumption of intrinsic periodicity implies that the Goldstone can vary only by finite quantities 

\section{Basic applications}

The classical-relativistic theory of elementary space-time cycles is not only completely equivalent to all the mains formulations of QM, it is also a powerful method to solve practical quantum phenomena.  
Here we will shortly report the basic procedures to describe and analytically solve, within the formalism of ECT, textbook problems of QM in a very straightforward way.  

\subsection{Non-Relativistic Free Particle}

We have proven that a quantum-relativistic free particle of energy $E$ is a ``periodic phenomenon'' constrained to have  persistent time periodicity $T = h / E$. In analogy with a particle in a periodic time box or a vibrating string, this constraint of periodicity leads to the harmonic energy spectrum $E_n = n E = n h / T $. As usual we assume $n \in \mathbb N$ --- the negative vibrational modes refer to antimatter. 

As already inferred in Sec.(\ref{states}), due to the controvariance of $T^\mu$, (\ref{disp:relat}), the time period $T$ is frame dependent.  The resulting dispersion relation of the energy spectrum (\ref{disp:rel:spectr}) is thus equivalent to that prescribed by ordinary second quantization (after normal ordering) for bosonic particles.  Similarly the spatial periodicity $\lambda$ implies the harmonic momentum spectrum $p_{(n) i}  = n h / \lambda^i$. 
The Compton periodicity, fixed by the mass according to Compton relation $T_C = h / M c^2$, determines, through Lorentz transformations, the space-time periodicity  and thus the quantized energy-momentum in that inertial reference frame.  

As well known from ordinary relativity, in the non-relativistic limit, \ie $|\vec p| \ll M/c $, the rest energy of the particle can be omitted as it forms an infinite energy gap: $E = \sqrt{M^2 c^4 + p^2 c^2} \sim M c^2 + \frac{p^2}{2 M}$, so that, at a classical level, the  dispersion relation for a free  non-relativistic particle is $E_{class}(\vec p) =  \frac{p^2}{2 M}$. Equivalently, the Compton periodicity, \ie the undulatory counterpart of the mass, can be neglected as it tends to zero in the non-relativistic limit, $T_C \rightarrow 0$.   The quantization of the particle dynamics, which is a direct consequence of the Compton periodicity, can be therefore neglected. This implies that the periodic phenomenon, \ie the elementary particle dynamics,  has continuous energy and momentum spectra in the non-relativistic limit. The quantization associated to the Compton clock is lost and we have ordinary non-relativistic physics. 

Furthermore, in ECT it is possible to show that $|\Phi (x)|$ is spatially localized within the Compton length $\lambda_C = T_C c$ along the classical particle path \cite{Dolce:2009ce}. To check this  it is sufficient to plot the modulo square of the wave-function (hint: in the wave phase neglect the term $M c^2$ coming from the non relativistic expansion of the dispersion relation). Thus, as  $\lambda_C \rightarrow 0$ we get a point-like description of the particle path in the non-relativistic limit (Dirac delta along the classical path). We have the ordinary non-relativistic classical path.

 In general, in ECT the corpuscular description is obtained in the limit of very small temporal periods --- \eg electromagnetic modes at very high frequency (Black Body UV region) can be described by photons, see Fig.(\ref{BlackBody}). The undulatory nature is predominant in the opposite limit, which coincides with the relativistic limit for massive particles.  ECT correctly describes the corpuscular counterpart of massless fields (\eg the electromagnetic radiation)  and the undulatory counterpart of massive particles. Actually we are just enforcing  undulatory dynamics into the space-time geometrodynamics.  Of course, this also yields an intuitive interpretation of, for instance, the double slit experiment, see early papers about ECT.

\subsection{ One dimensional Schr\"odinger problems: closed orbits}

In non-relativistic quantum problems $T_C$ can be neglected as already said. However, according to the postulate of ECT, elementary particles have an intrinsically cyclic nature, which becomes manifest again as soon as the particle is constrained in a potential. In the non-relativistic limit the intrinsic space-time periodicity of a particle, \ie the BCs quantizing the system, is determined by the potential $V(\vec x)$ rather than by the Compton periodicity. Our postulate is in fact that elementary particles are periodic phenomena. 

We immediately see  that the solution of many non-relativistic problems such as the particle in a box, potential well, the Dirac delta potentials, tunnel effect, and so on, is essentially identical to the ordinary ones: they in fact are easily solved by imposing the BCs prescribed by the potential, in full agreement with ECT.  

Similarly to the vibrating string, the energy and momentum spectra are determined by the requirement that, as prescribed by the postulate of intrinsic periodicity in ECT, the particle must have closed orbits along the phase-space of the potential. 
The condition of closed orbits is given by (\ref{BS:energymom}) or, by assuming a Morse factor, by (\ref{BS:morse}). In short, in non-relativistic problems the ``overdetermination'' of the system is provided by the potential. 

We have to calculate the duration of a orbit in the phase-space of the potential. Only those orbits in which the wave-length and the time period enter an integer number of recurrences are allowed for the quantum system. Thus the solution of simple non-relativistic Schr\"odinger problems is possible in perfect analogy to the Bohr-Sommerfeld quantization or to the WKB method. 

Once that the energy spectrum is known, the harmonics constituting the wave-packet of the system are derived by solving the equations of motion, in order to determine the local modulations of periodicity from the potential $V(x)$. As already said in the derivation of the axiom of motion from the postulate of intrinsic periodicity in Sec.(\ref{motion}), in the Hilbert space notation the evolution law of our periodic phenomenon in the potential $V(x)$ is given by the Schr\"odinger equation 
\begin{equation}\label{Schro:int}
i \hbar \frac{\partial}{\partial t} |\Phi(\vec x,t) \rangle = \mathcal H_{Class}(x) |\Phi( x,t) \rangle = \left(\frac{{\mathcal P}^2}{2 M} + V(x)\right)  |\Phi(x,t) \rangle \,. \end{equation}

 The Schr\"odinger equation encodes exactly the equations of motion expressed in Hilbert notation of all the harmonics constituting the ``periodic phenomenon''.    Below we will see some  examples about the validity of this approach.  We will obtain, of course, the same results reported in textbooks, but in a direct way. 

\subsection{Quantum harmonic oscillator, quantum anharmonic oscillator, linear potential, and the other textbook problems of quantum mechanics. }

The quantum harmonic oscillator of potential $V(x) = \frac{1}{2} m \omega^2 x^2$ can be easily solved by considering the pendulum isochronism: every orbit has the same fundamental period $T = 1 / f = 2 \pi / \omega$, independently on the libration.  This corresponds to the case of the ``periodic phenomenon'' with persistent time periodicity. We are in the case of an homogeneous string vibrating in time (\ref{BS:pers}). Thus the energy spectrum of the quantum harmonic oscillator is trivially harmonic $E_n = (n + \alpha) h f $ where we have assumed a generic Morse factor. According to (\ref{Schro:int}),  the spatial component of the n$^{th}$ harmonic of this ``periodic phenomenon'' is given by the same differential equation of the ordinary description 
\begin{equation}\label{QHO:schro}
-\frac{\hbar^2}{2 M}\frac{\partial^2 \phi_n(x)}{\partial^2 x} + \frac{M f}{4 \pi} \phi_n(x) = h f \left(n + \frac{1}{2}\right)\phi_n(x)\,,\end{equation}
where we have assumed $\alpha = 1 / 2 $ in order to have vanishing value (BCs) of the harmonics at the spatial infinite. These are the spatial BCs prescribed by the harmonic potential, in agreement with ECT. See below for the Coulomb potential.
  
We have the correct result also in non-trivial problems. The anharmonic potential is the harmonic potential plus a quartic term $\epsilon x^4 / l^4$ with $\epsilon \ll 1$ and $l = \sqrt{h / M f}$. In this case the requirement of close orbits prescribed by ECT yields a correction $\Delta E_n = \frac{3}{4} \epsilon (2 n^2 + 2 n)$ to the quantum harmonic oscillator energy spectrum. For the linear potential $V(x) = m g x$ (gravitational potential for small distances) the quantized energies satisfying closed orbits turn out to be $E_n = \frac{1}{2} [3 \pi (n + 1 / 4)]^{2/3} (\hbar M g^2)$. 

Notice that all these examples are commonly obtained in textbook through long and complicated calculations whereas in ECT they are straightforward. 
The other textbooks one-dimensional quantum problems such as the particle in a box, potential well, the Dirac delta potentials, tunnel effect, and so on, are directly solved by imposing spatial BCs, in full agreement with ECT.  

%Notice that all these examples are commonly obtained in text book through long and complicated calculations whereas in ECT they are straightforward.  

In short, \emph{in all the possible textbook one-dimensional quantum problems the postulate of intrinsic periodicity yields,  in a very straightforward way, exactly the same solutions known in literature}.

\subsection{Derivation of the second quantization: the creation and annihilation operators }\label{second:quant}

We have already described that the quantized spectrum inferred from the postulate of periodicity for the elementary relativistic particles is exactly that prescribed by second quantization, see also footnote.(\ref{foot}). 

The exact solution of the quantum harmonic oscillator allows us to introduce the Ladder operators in ETC, \ie the creation and annihilation operators of QFT. They are defined as functions of the position and momentum operators introduced in Sec.(\ref{observables}):  $a^\dagger = \sqrt\frac{M \omega}{ 2 \hbar} (x + \frac{i}{M \omega} \mathcal P) $  and $a = \sqrt\frac{M \omega}{ 2 \hbar} (x - \frac{i}{M \omega} \mathcal P)$. 

As well known from ordinary QFT, the commutation relation $[a^\dagger,a]=1$ is a direct mathematical implication of the ordinary commutations $ [x, \mathcal P] = i \hbar  $. The commutation relations of QM have been directly derived from the postulate of intrinsic periodicity. We can safely state  the commutation $[a^\dagger,a]=1$ are fully satisfied in ECT. From this it is possible to build the Fock space of QFT, as we will see in more detail. 

This represents the proof of the equivalence between ECT and the second quantization. Second quantization in fact is the quantization method of QFT based on commuting ladder operators of the quantum harmonic oscillator associated to every fundamental (frequency) mode of the classical field. Elementary cycles correctly describes the elementary quantum harmonic oscillator constituting every quantum fields, see \cite{Dolce:cycles;Dolce:FPIpaths} or similar papers.  

\subsection{Three dimensional Schr\"odinger problems, Coulomb potential, quantum numbers and Fock space}\label{Coulomb}

In spherical problems, beside the space-time periodicity considered so far,  it is necessary to consider the spherical periodicity as additional constraint ``overdetermining'' the relativistic dynamics. Indeed, similarly to the space-time periodicity of the periodic phenomenon (topology $\mathbb S^1$ determined by the single fundamental Compton periodicity), spherical problems are described by the two \textit{spherical angles} $\theta$ and $\varphi$ of (static) periodicities $\pi$ and $2 \pi$, respectively (topology $\mathbb S^2$). 

According to ordinary QM it is sufficient to impose  spherical periodicity as constraint (without any further quantization condition) to obtain the correct quantization of the corresponding physical conjugate variables, \ie of the quantization of the angular momentum of the system.  

We have seen that the space-time periodicities, Lorentz projections of the single Compton periodicity, implies a decomposition in harmonics $\phi_n(|\vec x|, t)$ denoted by the quantum number $\{n\}$ (one quantum number for the topology $\mathbb S^1$ of the Compton periodicity $T_C$). Similarly, the periodicity of the two spheric angles implies additional harmonic sets described by two corresponding quantum numbers $\{m,l\}$ (two quantum numbers for the two fundamental periodicities of $\mathbb S^2$). They are essentially the harmonics of a vibrating spherical membrane $Y_l^m(\theta, \varphi)$. They form a complete, orthogonal set defining a corresponding Hilbert space $\mathsf H_{sphere}$. 

The resulting solutions of the Schr\"odinger equation are the \textit{spherical harmonics}. The generic solution of a spherical problem is thus described by the three quantum numbers, corresponding to the harmonic eigenmodes $\phi_n(|\vec x|, t) Y_l^m(\theta, \varphi)$. For instance, to describe the three dimensional quantum harmonic oscillator we have to substitute the space-time harmonic solution $\phi_n(|\vec x|, t)$ with the solution of (\ref{QHO:schro}). 

%\begin{figure}\label{Fig1}
%\centerline{  \includegraphics[width=3cm]{membrana.pdf} ~~~~~ %\includegraphics[width=8cm]{spheric1.pdf} }
%\caption{\footnotesize  Representation of the spherical harmonics %$Y_l^m(\theta, \varphi)$, that are equivalent to the vibrational modes %of a spheric membrane. As proven by ETC, their combination with the %harmonics of the space-time vibrations of %he electron in a Coulomb %potential reproduces exactly the atomic orbitals.} %\abel{spheric}
%\end{figure}

Notice that in ECT the quantization by means of the constraint of intrinsic space-time periodicity is the space-time equivalent of the universally accepted and tested quantization of the angular momentum in terms of the intrinsic periodicity of the spherical angles --- is the relativistic generalization of ``the particle in a box''. Indeed no further quantization conditions for the angular momentum are required except  spherical periodicity (or its deformation). In other words, in ECT the space-time coordinates are treated as (non-independent) Minkowskian angular variables. This yields the quantization of the energy momentum exactly as the spheric angles yields the quantization of the angular momentum. \emph{The quantization of the angular momentum is another undeniable confirmation of the consistence of ECT}.

Another typical example of Schr\"odinger problem with spherical symmetry is the Hydrogen atom. The requirement of close space-time orbits for the electrons in a Coulomb potential  leads to the atomic energy levels $E_n = - \frac{13.6 eV}{n^2}$. Notice that, contrarily to  Bohr's original description, this does not  mean circular orbits. 
The requirement of closed space-time orbits, resulting from the Compton periodicity of the electron, is by far more general and fundamental with respect  to Bohr's model. The circular orbit is just a particular case. The only requirement is that the orbits must be closed in space-time, without any prescription about the shape. 

 In addition to the space-time periodicity, and the related principal number $n$,  the spherical symmetry of the Coulomb potential yields the additional spheric periodicity, resulting in the quantization of the angular momentum in terms of the quantum numbers $\{m,l\}$, with $0 \le l \le n - 1$ and $-l \le m \le l$. Thus the atomic orbitals can be regarded as the combination of vibrational modes  associated to the intrinsic space-time periodicity, modulated by the Coulomb potential, and the \textit{spherical harmonics} associated to the spherical symmetry. Notice that a careful application of the Bohr-Sommerfeld quantization, similarly to the method proposed here, yields the correct description of more advanced atomic physics such as atoms with more then one electrons, the Zeeman effect, and so on  \cite{refId0}. The zero-point energy is determined by the BCs at spatial infinity associated to the Coulomb potential. 

From a formal point of view, the composition of the space-time periodicity  (topology $\mathbb S^1$) and the spherical periodicity  (topology $\mathbb S^2$) is described by the product of the Hilbert spaces associated to these fundamental periodicities (topology $\mathbb S^1\otimes \mathbb S^2$): $\mathsf H_S = \mathsf H \otimes \mathsf H_{sphere}$ . The resulting Hilbert basis defined  by $\phi_n(|\vec x|, t) Y_l^m(\theta, \varphi)$ is represented in the Hilbert space by %$| \phi_n \rangle \otimes |Y_l^m \rangle$ or, with an improved notation,  
$| n,m,l \rangle = | n \rangle \otimes |m, l \rangle$ (three quantum numbers for the three fundamental periodicities of $\mathbb S^1\otimes \mathbb S^2$). This example shows that in ECT to every fundamental periodicity is associated a different quantum number. 

As anticipate in the previous sections, the composition of two elementary particles, \ie two distinguished space-time periodic phenomena, is thus given by  $| \Phi_1, \Phi_2 \rangle = | \Phi_1 \rangle \otimes | \Phi_2 \rangle $ with corresponding Hilbert basis  $|n_1, n_2 \rangle = |n_1\rangle \otimes |n_2 \rangle$.  If we consider the composition of many periodic phenomena we obtain the ordinary Fock space. In this way we can add the final formal element to the equivalence between ECT and relativistic QM: \emph{the composition of elementary intrinsic periodicities, as for instance multi-particle states, is consistently described by the Fock space in ECT.}

\subsection{Condensed matter and the role of the temperature}\label{temperature}

We shortly describe the role of the temperature in ECT  \cite{Dolce:2009ce,Dolce:Dice2014,Dolce:Dice2012}.

 Temperature is the manifestation of random collisions among particles (thermal noise). During these collisions, the ``periodic phenomena'' representing elementary particles have sudden jumps of periodic regimes, as consequence of the phase-harmony. 
 
So far we have consider single particles, which are ``periodic phenomena''.  A system of ``periodic phenomena'' at zero temperature (without collisions) tends to synchronize all the periodicities similar to Huygens' synchronization of clock's pendulums or metronomes \footnote{This is due to a residual electromagnetic interaction which, according to the geometrodynamical description of gauge theory in ECT, is the manifestation of peculiar local ``tunings'' of the particles space-time periodicities  \cite{Dolce:tune}.}. In the terminology of condensed matter we have that a system of particles at zero temperature condensates to ``complete coherence''. In this case the cyclic nature of QM is manifest as confirmed by all the experimental observations in condensed matter. This is the natural state of a system of elementary cycles.  
 
  However, the thermal noise is a  chaotic interference in the autocorrelation of the ``periodic phenomena'' constituting the system. Each particle has sudden variations of periodicities with characteristic time $\beta = \hbar / k_B \mathcal T$, being $k_B$ the Boltzmann constant and $\mathcal T$ the temperature. Thus the quantum system at finite temperature is characterized by Poissonian decay of the ``complete coherence'' of the particles with a characteristic (reduced) thermal time $\beta = h / k_B \mathcal T  $. The result is a frequency damping $e^{-\hbar \omega / k_B \mathcal T} = e^{- 2 \pi \beta / T}$ of the cyclic behavior of  periodicity $T$, see below.  
  
  In other words, while the Minkowskian periodicity $T$ of quantum mechanics tends to form perfect coherent states (``periodic phenomena''), the thermal time $\beta$ describes a damping associated to the thermal noise which tends to destroy the perfect recurrences of the pure quantum systems. Thus, if the $T \ll \beta$ the system can autocorrelate (condensate) and give rise to pure quantum phenomena  whereas in the opposite limit the thermal noise breaks the quantum recurrences before they can autocorrelate, leading to the classical  behavior (\eg ordinary electric resistance or other aperiodic behaviors).  
  
  \subsubsection{Matsubara theory}
  
  The damping factor for the $n^{th}$ harmonics of the periodic phenomenon  is $e^{-\hbar \omega_n / k_B \mathcal T} = e^{-2 \pi n \beta / T}$ directly related to the Boltzmann statistics. Indeed it describes the probability to populate the n$^{th}$ vibrational mode of the system.  It can be equivalently written as $e^{- n \omega \beta / \hbar}$. For instance, in a Black Body at temperature $\mathcal T$ many vibrational modes are populated in the IR region whereas only few modes are accessible in the UV region, see Fig.(\ref{BlackBody}). 
  
    The frequency damping factor can be equivalently written as $e^{- n \omega \beta }$. By comparison  with the typical harmonic expansion of the ``periodic phenomenon'' $e^{- i n \omega t}$ (Minkowskian periodicity), we see that $\beta$ describes an Euclidean ``periodicity'' ($t \rightarrow - i t$). We have found the fundamental link between ECT and the Matzubara they.
     
    In other words, in the Matzubara theory the quantization of a statistical system at temperature $\mathcal T$ is obtained by imposing intrinsic  Euclidean ``periodicity'' $\beta$ exactly as in ECT the quantization of relativistic dynamics of energy $E$ is obtained by imposing intrinsic Minkowskian periodicity $T$. Again we have a striking confirmation that quantum phenomena, in this case of thermal system, are manifestations of intrinsic periodicity. This is another success of ECT. It proofs for the first time the physical meaning of the so-called ``mathematical trick''  ( \ie a mathematical method that so far has been considered to have no fundamental physical meaning) of the Euclidean periodicity in the Matzubara theory.
    
    We also see that the Euclidean periodicity has an opposite meaning with respect to the periodicity in the Minkowskian time $t$. The latter is a real persistent periodicity (exponential of imaginary numbers), the former is an exponential decay (exponential of real numbers), \ie a dumping of the persistent periodicity. ECT thus explains the opposite physical role of the Minkowskian time and the Euclidean time (which are often confused by some authors, because its physical meaning as been so far obscure), as well as the meaning of the Wick rotation. 
  
   Thus, at very low temperature $T \ll \beta$ only the fundamental vibrational mode $n=1$ will be populated, whereas at high temperature many vibrational modes must be considered, so that the spectrum can be approximated by a continuum (see the interpretation of the classical limit of the Black Body radiation or of the Bohr atom). Surprisingly, these simple considerations are sufficient to describe, even formally, the most fundamental phenomena of condensed matter, such as superconductivity or graphene physics, as we will see below  \cite{Dolce:SuperC,Dolce:EPJP,Dolce:Dice2014}

\subsubsection{Dirac quantization for magnetic monopoles}

In ECT gauge interactions, such as the electromagnetic interaction, can be directly inferred (without postulating it) from the  space-time geometrodynamics in deep analogy with the description of gravitational interaction in general relativity \cite{Dolce:tune}, see next section.  
For the scope of this paper it is nevertheless sufficient to introduce electromagnetic interactions in the usual way, \ie by assuming the minimal substitution $p'_\mu(x) = p_\mu - e A_\mu(x)$ where $A_\mu(x)$ is the electromagnetic potential. 
With this substitution in the modulated periodic phenomenon, \eg in (\ref{modul:solution}), in analogy with the Bohr-Sommerfeld quantization, the condition of intrinsic periodicity leads directly to the \emph{Dirac quantization for magnetic monopoles} (and to a link to \emph{Dirac strings})
\begin{equation}e^{\frac{i e}{\hbar} \oint_{T^\mu(X)} A_{\mu} d x^\mu} = e^{-i 2 \pi n} ~~~~\rightarrow ~~~  \oint_{T(X)} A_\mu d x^\mu = 2 \pi n \hbar \,.\label{Dirac:string}\end{equation}

\subsubsection{Superconductivity and graphene physics}

The Dirac quantization of magnetic monopoles obtained above is at the base of the derivation of superconductivity and its fundamental phenomenology in ECT, directly from first principles of QM rather than from empirical models such as BCS theory  \cite{Dolce:SuperC}. Indeed, if we consider the gauge invariance $\Phi(\vec x, t) = \mathsf{U} ( \vec x, t) \Phi(\vec x, t)$ where $ \mathsf U( \vec x, t) = e^{-i \frac{e}{\hbar c} \theta (\vec x, t)}$, the condition of intrinsic periodicity in this case implies that the Goldstone $\theta (\vec x, t)$ is periodic and defined modulo factors $2 \pi n$:
\begin{equation}
\frac{e}{\hbar c} \theta (\vec x, t) = \frac{e}{\hbar c} \theta (\vec x, t + T) + 2 \pi n  \,.
\end{equation} 
From the Stokes theorem and by considering a contour on an electric conductor in which the field is in pure gauge $A_\mu = \partial_\mu \theta$ we find that the magnetic flux through the area $S_\Sigma$  is quantized 
\begin{equation}
\int_{S_\Sigma} \vec B \cdot d \vec S = \oint_\Sigma \vec A \cdot d \vec x =  \oint_\Sigma \cdot d \vec x = n \frac{h c }{e}\,.
\end{equation} 

As the magnetic flux is quantized, the current cannot smoothly decay, so that we have superconductivity  \cite{Weinberg:2014ewa}. Similarly it is possible to derive the other effects characterizing superconductivity such as the Meissner effect, the Josephson effect, the Little-Parks effect, energy gap opening, and  and so on, see  \cite{Dolce:SuperC,Dolce:EPJP,Dolce:Dice2014}. From here, by following the line of \cite{Weinberg:2014ewa}, it is possible to fully derive the BCS theory. 

%Here we will introduce the role of the temperature in the EC theory, which can be directly generalized to CA. In CA, as in undulatory mechanics and EC theory, the time period and the energy are ``two faces of the same coin''. Therefore interactions, \ie variations of energy $\hbar \omega = 2 \pi \hbar / T$, correspond to local modulations of the period $T$. 

Remarkably, the cyclic dynamics characterizing ECT, the relativistic modulation of periodicity and the related dispersion relation of the energy spectrum, the description of antimatter as negative vibrational modes, and so on, can be directly tested in carbon nanotubes. Actually we have found  \cite{Dolce:SuperC,Dolce:EPJP,Dolce:Dice2014} that ECT provides an elegant and simple new technique to derive the essential electronic properties of carbon nanotubes or similar graphene systems.

Due to the peculiar graphene lattice electrons behaves as massless particles in a two dimensional space-time. That is, the temporal and spatial recurrences of electrons in graphene are each other proportional. When a graphene dimension is curled up (compactification on a circle) to form a carbon nanotube the electrons acquire an effective rest mass (finite Compton periodicity, the electron at rest with respect to the axial direction have a residual cyclic motion) and the others electronic properties exactly as predicted by ECT.

Summarizing the result of \cite{Dolce:SuperC,Dolce:EPJP,Dolce:Dice2014} we have that graphene physics is a direct experimental confirmation of ECT. In nanotubes the Compton periodicity is rescaled to time scales accessible to modern timekeepers, allowing us to ideate interesting experiments, and indirectly investigate the cyclic dynamics beyond QM.

\section{Implications in modern physics}

In this paper we have limited our introduction to ECT to basic aspects of quantum theory, typically reported in QM textbooks. It must be however said that ECT has amazing applications and confirmations in advanced aspects of modern theoretical physics.

\subsection{Gauge interactions from space-time geometrodynamics}

In ECT both gravitational and gauge interactions are derived from space-time geometrodynamics, respectively as local deformations of the flat space-time metric and of local rotations of the space-time boundary  \cite{Dolce:tune}. The latter is further success of ECT which has no comparison with other unification attempt, though the road  was essentially that already suggested by eminent physicists such as Einstein, Weyl, Kaluza and so on (the missing ingredient in all these cases was the hypothesis of intrinsic periodicity). 

ECT postulates that elementary particles are relativistic reference clocks.  We have seen that the local variations of energy-momentum $p_\mu (x)$ during interactions are in fact equivalently described by local and retarded modulations of their de Broglie-Planck periods $T^\mu(x)$. In perfect agreement with general relativity the modulations of elementary clock rates in different space-time points  are encoded in the corresponding curved space-time metric. 

It is interesting to remember that Einstein's original approach to general relativity was in terms of clocks modulations. Actually it is possible to derive the whole construction of general relativity in terms of clocks modulations and phase harmony, as also reported in literature, see \cite{Dolce:tune} and citations thereby.  The modulations of clock rate prescribed by the phase-harmony, and in turn by ECT, are in perfect agreement with those prescribed by general relativity. 

In short, see \cite{Dolce:cycles,Dolce:tune,Dolce:AdSCFT,Dolce:FQXi,Dolce:Dice,Dolce:2009ce} for the full derivation, let us consider a particle in a Newtonian gravitational potential $V(|\vec x|) = - G M_{\bigodot}/|\vec x|$ of a planet of mass $M_{\bigodot}$, where $G$ is the Newton constant. Its energy varies with respect to the free case varies as $E \rightarrow E'(|\vec x|)= (1 + G M_{\bigodot}/|\vec x|)E$. According to the phase harmony $E = h / T$ the rate of the particle clock turns out to be locally modulated with respect to the free case  as $T \rightarrow T'= T/(1 + G M_{\bigodot}/|\vec x|)$. Therefore the particle clock, similar to Einstein's clocks, runs slower inside the gravitational well. This immediately means that the two fundamental aspects of general relativity, such as time dilatation and gravitational red-shift, can be directly inferred from the postulate of intrinsic periodicity. 

Similarly we have the correct description of the wave-length modulations in the gravitational well. This implies that the modulation of space-time periodicity prescribed by ECT for a weak Newtonian potential is encoded by the ordinary Schwarzschild metric. By following this line, and considering also self gravitational interaction, as also know in literature, it is possible to fully derive the Einstein equation. ECT is in perfect agreement with general relativity, it actually represents an intuitive way to infer fundamental aspects of Einstein theory \cite{Dolce:cycles,Dolce:ECunif,Dolce:TM2012,Dolce:Dice2012,Dolce:cycle,Dolce:ICHEP2012,Dolce:tune,Dolce:AdSCFT,Dolce:FQXi,Dolce:Dice,Dolce:2010ij,Dolce:2010zz,Dolce:2009ce}.

The intrinsic periodicity of elementary particles, also implicitly in undulatory mechanics, is realized by defining the corresponding relativistic action in compact space-time dimensions, with local four-compactification length $T^\mu(x)$ and covariant local PBCs ---  $T^\mu(x)$ is loaclly determined by the kinematical state of the particle. In other words, the boundary of the action must vary locally and contravariantly encoding interactions.  We find that interactions are encoded in the geometrodynamics of the boundaries, providing a fundamental justification to \textit{Holography} \cite{Dolce:tune,Dolce:AdSCFT,Dolce:cycles}.  

The local deformation of the space-time metric of the elementary clocks leads to gravitational interaction as we have seen. In addition to this, ECT allows another peculiar mechanism to modulate the clock rates: due to the intrinsically compact nature of space-time implies that even local rotations of the space-time boundary can generate local modulations of Broglie-Planck period. In turn the local rotations of the boundary describe particular kind of interactions (this is a peculiar property of ECT\footnote{These transformations are related to the Killing vectors which in ordinary relativistic theories defined in non-compact space-time have no physical effects.}). These particular kind of interactions cannot be gravitational ones, because the boundary can rotate leaving the metric flat.

 It turns out that these particular interactions are formally equivalent in all the fundamental aspects to ordinary gauge interactions, \eg electromagnetism. In simple words: \emph{we are able to infer all the fundamental aspects of gauge interactions, without postulating gauge invariance, directly from the assumption of intrinsic periodicity and related space-time geometrodynamics}. Again, the constraint of periodicity represents the quantization condition ``overdetermining'' the gauge dynamics. \emph{The result is a formal equivalence to ordinary QED}.  \cite{Dolce:tune} 
 We have an unedited unified geometrodynamical view of both gauge and gravitational interactions. 

\subsection{Correspondence with extra-dimensional theories and string theory}

ECT inherits fundamental aspects of modern theories, see   \cite{Dolce:AdSCFT}  and citations thereby. From a mathematical point of view the cyclic (or, more in general, compact) world-line parameter associated to the Compton periodicity of a elementary particle,  plays a role very similar to the cyclic (or compact) extra-dimension of the \textit{Kaluza-Klein theory} --- surprisingly the geometrodynamical dynamical description of gauge interactions in this analogy proves the so called Kaluza's miracle. 

For these reasons, the world-line parameter of ECT is also named ``virtual extra dimension''. The combination of the correspondence of ECT, on one hand to classical extra-dimensional theories, and on the other hand to ordinary QM (Feynman path integral), leads to an intuitive formal derivation of \textit{Maldacena's conjecture} (also known as \textit{AdS/CFT} or \textit{gauge/gravity duality})  by means of simple semi-classical arguments  \cite{Dolce:AdSCFT,Dolce:ICHEP2012}.
As clearly stated, for instance, by Witten the \textit{AdS/CFT} correspondence is a classical geometry to quantum behavior correspondence. 

 The Compton periodicity implies that the world-line parameter (or the proper time) of the quantum particle must be intrinsically cyclic with periodicity $T_C$. This plays the role of the cyclic (compact) world-sheet parameter of ordinary \textit{closed strings} (open strings). In this way ECT inherits fundamental good behaviors  of ordinary string theory, without inheriting the problematic aspects such as the necessity to introduce extra-dimensions; the target space is the purely four-dimensional space of ECT without inconsistencies. \emph{ECT must be regarded as a one dimensional string theory defined in a cyclic world-line and target purely four-dimensional compact space-time.}

\subsection{ Time cycles and the interpretation of time flow }\label{timeflow}

The concept of time is probably the most mysterious concept of physics. In ECT introduces an extraordinary novel element, with respect to ordinary interpretations, to successfully address the problem of time in physics. It clearly defines the elementary clocks of Nature. Massive elementary particles are the elementary clocks of Nature whose rates are determined by the masses. %With this we this new element at hand we can work out the meaning of time in physics.   

In other words, ECT introduces a cyclic character to the ordinary Minkowskian time. The cyclic nature of time must not surprise: it has been  supported by some of the most notable philosophers from the beginning of the human civilization up to now (but curiously ignored by modern physicists, despite the fact that the universe is full of effective, isolated periodic phenomena). 

To fully understand the concept of time in this approach it is necessary to distinguish between the cyclic time coordinate and the time flow. The former refers to the Minkowskian time of relativity which is cyclic in ECT. It is typical of pure quantum phenomena and it is the inedited aspect introduced by our approach. The latter involves the thermal time (Euclidean periodicity). It's origin is statistical. As already argued, its role is clarified by ECT with respect to ordinary interpretations of the interplay between QM and thermodynamics.   

We have postulated that every particle, \ie every elementary constituent of nature, is a reference clock whose rate is determined by the particle's mass  \cite{Lan01022013}. This means that the Minkowskian time coordinate has a cyclic (angular) nature for elementary particles. 
A particular case is given by massless particles (photons and gravitons) whose internal Compton clock is ``frozen'' --- ECT must not be confused with cyclic cosmology: roughly speaking, in ETC every elementary particle can be regarded as an ultra-fast cyclic universe. 

This radically new description of relativistic time demands for a radical reconsideration of the QM paradoxes. Indeed, we have largely proven that such relativistic cyclic dynamics are formally identical in all the fundamental aspects to relativistic QM. Using Einstein's terminology, QM is directly inferred from the ``overdetermination'' of relativistic dynamics. 

Elementary particles are ``periodic phenomena''. This intrinsic periodicity is manifest in pure quantum systems such at systems at zero temperature, or ultra-relativistic free particles.  On the other hand we must remember that interactions correspond to local and retarded modulations of the periodicities of these elementary clocks. Summarizing this implies the following fundamental aspects attributed to time , \cite{Dolce:2009ce,Dolce:FQXi,Dolce:cycle,Dolce:TM2012}. Time ordering:  interaction is an event in space-time from which it can be established a past and a future, as it determines a variation of the particles periodic regimes.  Causality: the periodicity is locally determined by the energy and in turns by the ordinary retarded relativistic potentials. Ergodic evolutions: a simple system of two or more non interacting ``periodic phenomena'' is an ergodic (aperiodic) system. Chaotic evolution: at finite temperature, due to the chaotic interactions (the collision determining the thermal noise) among ``periodic phenomena'' the system will have chaotic evolution described by the statistical laws, and thus by laws of thermodynamics (including inreversibility and the concept of Entropy). In short the arrow of time emerges in a relational, statistical, thermodynamical way by combining  the modulated ticks of all the elementary clocks of Nature constituting every macroscopic system.

\section{Conclusions}

Where is and what is fixed at the boundary of relativistic space-time? Why is time so peculiar with respect to space? What does QM tell us about the nature of time? What is the intimate relationship between relativity and QM, or between gravitational and gauge interactions? 
 These fundamental questions cannot be answered by simply invoking relativistic theory and the ordinary properties of Minkowskian time. 
 
We have largely proven, with incontestable demonstrations \cite{Dolce:cycles,Dolce:ECunif,Dolce:FPIpaths,Dolce:SuperC,Dolce:TM2012,Dolce:Dice2012,Dolce:cycle,Dolce:ICHEP2012,Dolce:tune,Dolce:AdSCFT,Dolce:FQXi,Dolce:Dice,Dolce:2010ij,Dolce:2010zz,Dolce:2009ce,Dolce:EPJP,Dolce:SuperC,Dolce:Dice2014}, that the ``missing link'' to answer these questions is the postulate of ``intrinsic periodicity'' of elementary particles. The wave-particle duality must be encoded directly into the fabric of relativistic space-time. In particular, \emph{the Minkowskian time of relativistic theories must have an intrinsically cyclic (or, more in general, compact) nature.} 

The postulate of intrinsic periodicity is implicit in the combination of  Newton's law of inertia and undulatory mechanics. As also suggested by de Broglie, every elementary particle in Nature is a ``periodic phenomenon''.  Inevitably, every system in nature must be described in terms of elementary space-time cycles. All in all the line was already drawn by Einstein himself: the condition of periodicity can be imposed as constraint to ``overdetermine'' relativistic mechanics.  With this element at hand  we obtain exactly and rigorously the unification of relativistic and quantum mechanics. 

The mathematical demonstrations are incontrovertible: under the postulate of intrinsic periodicity, quantum mechanics emerge as novel relativistic phenomena, without any explicit quantization condition. The resulting cyclic dynamics are mathematically equivalent (without fine-tunings or hidden variables) to ordinary relativistic quantum mechanics in all its main formulations, such as the canonical formulation, the Dirac quantization, the second quantization of quantum field theory, and the Feynman path integral formulation, as well as in all the fundamental phenomenology of quantum world. QM emerge as statistical, effective description of the ultra-fast cyclic dynamics associated to elementary particles, whose time scales are still beyond the temporal resolution of modern timekeepers.  ECT also implies an unedited formulation of gauge interactions in terms of peculiar space-time geometrodynamics, absolutely parallel to the gravitational ones of general relativity.         

In this paper we have introduced --- in a pedagogical way --- the main quantum aspects of the ECT, giving  step by step demonstrations of some of its main results, as well as practical applications of this astonishing unified description of physics.

%
%\bibliographystyle{ws-rv-van}
%\bibliography{../comp3+1}
%\bibliography{../cycles}

\end{document}